\documentclass[usegraphicx,usenatbib,useAMS]{mn2e}\usepackage{times}
\usepackage{graphicx}
\usepackage{amssymb}
\usepackage{caption}
\usepackage{subcaption}
\let\url\relax
\usepackage{amsmath}
\usepackage{hyperref}
\DeclareMathAlphabet{\mathbi}{\encodingdefault}{\rmdefault}{\bfdefault}{\itdefault}
\DeclareRobustCommand{\bit}[1]{\ifmmode\mathbi{#1}\else\textbf{\textit{#1}}\fi}
\let\bolditalic=\bit
\def\apj{{ApJ}}
\def\apjs{{ApJs}} 
\def\apjl{{ApJL}}
\def\aap{{A.\&A}}
\def\aj{{AJ}}

\def\prd{{Phys Rev D}}
\def\nat{{Nature}}

\def\MNRAS{{MNRAS}}

\def\mnras{{MNRAS}}
\def\apj{{ApJ}}
\def\apjs{{ApJs}} 
\def\apjl{{ApJL}}
\def\aap{{A.\&A}}
\def\aj{{AJ}}

\def\jcap{{JCAP}}
\def\prd{{Phys Rev D}}
\def\nat{{Nature}}

\def\MNRAS{{MNRAS}}

\def\mnras{{MNRAS}}
\newcommand{\be}{\begin{equation}}
\newcommand{\ba}{\begin{eqnarray}}
\newcommand{\ee}{\end{equation}}
\newcommand{\ea}{\end{eqnarray}}
   
\def\vect#1{{\vec{\bolditalic #1}}}
\def\thetavec{\vect{\theta}}

\begin{document}

\title [The Jubilee ISW Project I]{The Jubilee ISW Project I: simulated ISW and weak lensing maps and initial power spectra results}
 
\author[W. A. Watson et al.]{W.~A.~Watson$^1$,\thanks{e-mail: 
W.Watson@sussex.ac.uk} J.~M. Diego$^{2}$, S.~Gottl\"ober$^3$, I.~T.~Iliev$^1$, A.~Knebe$^4$, \newauthor E.~Mart\'inez-Gonz\'alez$^{2}$, G.~Yepes$^4$, R.~B.~Barreiro$^2$, J.~Gonz\'alez-Nuevo$^2$, \newauthor S.~Hotchkiss$^5$, A.~Marcos-Caballero$^2$, S.~Nadathur$^6$, P.~Vielva$^2$
\\
$^1$ Astronomy Centre, Department of Physics \& Astronomy, Pevensey II 
Building, University of Sussex, Falmer, Brighton, BN1 9QH, United Kingdom\\
$^2$ IFCA, Instituto de F\'\i sica de Cantabria (UC-CSIC). Avda. Los Castros s/n. 39005 Santander, Spain.\\
$^3$ Leibniz-Institute for Astrophysics, An der Sternwarte 16, 14482 
Potsdam, Germany\\
$^4$ Departamento de F\'isica Te\'orica, Modulo C-XI, Facultad de Ciencias, 
Universidad Aut\'onoma de Madrid, 28049 Cantoblanco, Madrid, Spain\\
$^5$ Department of Physics, University of Helsinki and Helsinki Institute of Physics, P.O. Box 64, FIN-00014 University of Helsinki, Finland\\
$^6$ Fakult\"at f\"ur Physik, Universit\"at Bielefeld, Postfach 100131, D-33501 Bielefeld, Germany\\
}

\date{\today} \pubyear{2013} \volume{000}
\pagerange{1} \twocolumn \maketitle
\label{firstpage}

\begin{abstract}
We present initial results from the Jubilee ISW project, which models the expected $\Lambda$CDM Integrated Sachs-Wolfe (ISW) effect in the Jubilee simulation. The simulation volume is $(6~h^{-1}\mathrm{Gpc})^3$, allowing power on very large-scales to be incorporated into the calculation. Haloes are resolved down to a mass of $1.5\times10^{12}~h^{-1}\mathrm{M}_\odot$, which allows us to derive a catalogue of mock Luminous Red Galaxies (LRGs) for cross-correlation analysis with the ISW signal. We find the ISW effect observed on a projected sky to grow stronger at late times with the evolution of the ISW power spectrum matching expectations from linear theory. Maps of the gravitational lensing effect, including the convergence and deflection fields, are calculated using the same potential as for the ISW. We calculate the redshift dependence of the ISW-LRG cross-correlation signal for a full sky survey with no noise considerations. For  $\ell < 30$, the signal is strongest for lower redshift bins ($z \sim 0.2$ to $0.5$), whereas for $\ell > 30$ the signal is best observed with surveys covering $z \sim 0.6-1.0$.
\end{abstract}

\begin{keywords}
cosmology: cosmic microwave background---dark energy---large-scale structure of Universe---methods: numerical
\end{keywords}

\section{Introduction}
\label{intro:sect}

The recent results from the \emph{Planck} satellite \citep{2013arXiv1303.5062P} have shown the standard $\Lambda$ Cold Dark Matter ($\Lambda$CDM) cosmological model to be in good health. The universe, as we currently understand it, consists mainly of some form of dark energy or cosmological constant ($\Lambda$) and a cold dark matter component. The key challenges in cosmology, however, remain the same: we still need to uncover the secrets of the dark sector. What is dark matter? What are the properties of dark energy? 

To answer the latter question, the late-time integrated Sachs-Wolfe (ISW) effect \citep{1967ApJ...147...73S,1968Natur.217..511R,1994PhRvD..50..627H} can be a useful cosmological probe, since it is sensitive to the dynamical effects of dark energy and may thus be used to discriminate between different cosmological models \citep{1996PhRvL..76..575C, 2004PhRvD..69h3524A}. The effect is manifested as secondary anisotropies in the cosmic microwave background (CMB) radiation temperature, which are created when photons from the last scattering surface travel through time-evolving fluctuations in the gravitational potential, $\Phi$, caused by large-scale structure (LSS) along their paths. For a flat universe filled entirely with a pressureless fluid such as dark matter, at linear order $\Phi$ is constant with time, so that to first order the linear ISW effect is zero, although second order effects would arise, primarily due to the velocity field of the structures that seed the potential. The time evolution of $\Phi$ requires a significant non-pressureless component of the cosmological fluid \citep{1967ApJ...147...73S} or non-zero curvature \citep{1994ApJ...432....7K}. Given that \emph{Planck} shows the universe to be very close to flat \citep{2013arXiv1303.5076P}, a detection of the ISW effect constitutes a direct measure of the effects of dark energy.

However, the detection of the ISW effect is complicated by two factors. The first is that the amplitude of the effect on observationally relevant scales is an order of magnitude smaller than primordial anisotropies in the CMB. The second is that the ISW contribution to the CMB temperature power spectrum is greatest on large angular scales. This means that the measurement is very susceptible to cosmic variance and also that the detection of the signal through cross-correlation of CMB temperatures with LSS requires the use of galaxy surveys covering a large sky fraction and containing a very large number of galaxies \citep{2004PhRvD..69h3524A, 2008A&A...485..395D}.

Following the earliest reported detections by \citet*{2003ApJ...597L..89F, 2004Natur.427...45B, 2004PhRvD..69h3524A, 2004ApJ...608...10N}, most studies of the ISW effect have been based on a full cross-correlation between the CMB and different LSS catalogues that trace the matter density. Different techniques to achieve this calculate the cross-correlation in either real \citep[e.g.][]{2002PhRvL..88b1302B,2008PhRvD..77l3520G}, harmonic \citep[e.g.][]{2004PhRvD..69h3524A,2012MNRAS.427.3044S} or wavelet \citep[e.g.][]{2006MNRAS.365..891V,2007MNRAS.376.1211M} space. The results of these studies have been mixed, with reported detection significances ranging from low significance to $4\sigma$ \citep[see][for a recent study and a brief review of previous results]{2013arXiv1303.5079P}. Recently, the \emph{Planck} collaboration has also been able to cross-correlate the CMB map with a map of the reconstructed lensing potential, finding a $\sim2.5\sigma$ significant detection of the ISW-lensing cross-correlation \citep{2013arXiv1303.5079P}. \emph{Planck} has also obtained evidence for the ISW effect through a measurement (via the bispectrum) of the non-Gaussianity imprinted in the CMB due to this ISW-lensing correlation \citep{2013arXiv1303.5084P}.

A different approach using a stacking analysis of CMB patches along lines of sight that correspond to individual over- or underdensities identified in a galaxy survey was found by \citet*{2008ApJ...683L..99G} to give a detection significance $>4\sigma$, a result recently confirmed by \citet{2013arXiv1303.5079P} using the same lines of sight. The amplitude of the signal observed in this approach is, however, too large for the standard $\Lambda$CDM cosmology \citep*{Hunt:2008wp, 2012JCAP...06..042N, 2013JCAP...02..013F, 2012arXiv1212.1174H} and is currently unexplained. Subsequent stacking investigations using a different catalogue of voids have not shown the same strength of signal \citep*{2013arXiv1301.5849I, 2013arXiv1303.5079P}, adding to the mystery. Given the wide range of results and the uncertainties involved in their interpretations, a great deal of importance is placed on improving our theoretical understanding of the expected ISW effect in a $\Lambda$CDM cosmology. This may be best addressed by using large $N$-body simulations. 

Whilst the large-scale ISW effect is governed by the dark energy-driven time variability of the gravitational potential -- and is therefore observed in the radial direction -- variations in the tangential direction of the potential results in achromatic path distortions of the photons (i.e with no gain or loss of energy). These tangential distortions are the gravitational lensing effect \citep[see][for a review]{2008ARNPS..58...99H}. Lensing distortions concentrate on the small scales (of the order of a few arcminutes) and hence complement the large-scale ISW effect. The lensing effect does not depend (at least not to first order) on dark energy but is very sensitive to the distribution of the total mass. Due to this direct dependency on dark matter, gravitational lensing can produce reliable estimates of the matter power spectrum and thus provide independent and robust estimates of the cosmological model. Measurements of the CMB lensing effect \citep[for example, see ][ for results from \emph{Planck}]{2013arXiv1303.5077P} can be used to set constraints on the spatial curvature, dark energy or neutrino masses \citep{2013MNRAS.tmp.1228M} that are normally degenerate when only the CMB power spectrum is available. Gravitational lensing will be a source of confusion noise in future CMB polarization missions (like the proposed PRISM\footnote{\url{www.prism-mission.org}} mission) as the effect introduces B-modes from the primordial E-modes. Large simulations are needed to properly account for this source of systematic error and study ways of reducing its impact. Among these projects, future space missions such as Euclid \citep{2012SPIE.8442E..0ZA} will need to rely on realistic simulations that include not only the lensing effect due to large-scale structure but also the associated catalogs that trace that matter. 

Simulations will be needed to validate the methods employed in these future missions and much work has already been undertaken on the topic of lensing in this field \citep[see, for example,][]{1999MNRAS.310..453B,2000ApJ...530..547J,2003ApJ...592..699V,2008MNRAS.388.1618C,2008ApJ...682....1D,2008MNRAS.391..435F,2009A&A...499...31H,2009A&A...497..335T,2010ApJ...713.1322L,2011MNRAS.414.2235K,Carbone2013}. However, simulations are typically based on boxes that are much smaller than the Hubble volume (typically with $L_{box}\sim500~h^{-1}\mathrm{Mpc}$ to $1~h^{-1}\mathrm{Gpc}$, although \cite{2009A&A...497..335T} and \cite{2008MNRAS.391..435F} consider boxes of length $2$ and $3~h^{-1}\mathrm{Gpc}$ respectively). For future surveys a much larger volume would be more suitable especially for the case of CMB lensing where the lensing cross section peaks at around $z=1$ (i.e.\ around $2.3~h^{-1}\mathrm{Gpc}$).

To study both the ISW and weak lensing effects we have performed a large $N$-body simulation: the Juropa Hubble Volume, `Jubilee', simulation\footnote{\url{http://jubilee-project.org}} \citep{2013arXiv1305.1976W}. The simulation contains $6000^3$ particles in a box of side $6~h^{-1}$Gpc. It is therefore possible to use the simulation to model the ISW effect due to large-scale structure out to $z=1.4$ without having to repeat the box \citep[a shortcoming of previous, smaller, ISW simulations; see, for example,][and the discussion in \S~\ref{sect:isw_pow}, below]{2010MNRAS.407..201C}. Furthermore, with its high particle count we are able to directly resolve dark matter haloes that contain Luminous Red Galaxies (LRGs). This allows us to measure the cross-correlation between the simulated ISW and the large-scale structure traced by the LRGs on larger scales than has hitherto been possible. Direct measurement of the expected stacking signal from LRGs is also possible, as well as studies of the ISW-lensing cross-correlation.

This paper details methodologies for the creation of mock LRGs, all-sky weak lensing maps, and the ISW effect. It also presents initial results for the ISW-LSS cross-correlation signal. The results presented in this work relate to the pure ISW-LSS signal, with no signal-to-noise considerations. This paper is laid out as follows. We first detail the particulars of the Jubilee simulation in \S~\ref{sim:sect}, then provide an overview of how the ISW maps, LRG catalogues and weak lensing maps were created In \S~\ref{sect:meth}. We then present the results from these modelling procedures followed by the ISW-LSS cross-correlation signal in \S~\ref{sect:res}. Finally, in \S~\ref{sect:sum}, we conclude with some general comments on the implications of this work for future ISW-detection efforts, and briefly lay out the work we will be presenting on this topic in the future.

\section{The Jubilee simulation}
\label{sim:sect}

The results presented in this work are based on a large-scale structure $N$-body simulation, detailed in \cite{2013arXiv1305.1976W}. The simulation has $6000^3$ (216 billion) particles in a volume of  $(6~h^{-1}\mathrm{Gpc})^3$. The particle mass is $7.49\times10^{10}~h^{-1}\mathrm{M}_{\odot}$, yielding a minimum resolved halo mass (with 20 particles) of $1.49\times10^{12}~h^{-1}\mathrm{M}_{\odot}$, corresponding to galaxies slightly more massive than the Milky Way. LRGs ($\mathrm{M}_{\mathrm{halo}}\sim10^{13}~h^{-1}\mathrm{M}_{\odot}$) are resolved with $\sim100$ particles, and galaxy clusters ($\mathrm{M}_{\mathrm{halo}}>10^{14}~h^{-1}\mathrm{M}_{\odot}$) are resolved with $10^3$ particles or more. The simulation and most analyses were performed on the Juropa supercomputer at J\"ulich Supercomputing Centre in Germany (17,664 cores, 53 TB RAM, 207 TFlops peak performance) and required approximately 1.5 million core-hours to complete. The simulation was run on 8,000 computing cores (1,000 MPI processes, each with 8 OpenMP threads) using the {\small CUBEP$^3$M} $N$-body code, a P$^3$M (particle-particle-particle-mesh) code \citep{HarnoisDeraps:2012he}. {\small CUBEP$^3$M} calculates the long-range gravity forces on a 2-level mesh and short-range forces exactly, by direct summation over local particles. The code is massively-parallel, using hybrid (combining MPI and OpenMP) parallelization and has been shown to scale well up to tens of thousands of computing cores \citep[see][for a complete code description and tests]{HarnoisDeraps:2012he}.

We base our simulation on the 5-year WMAP results \citep{Dunkley:2008ie, Komatsu:2008hk}. The cosmology used was the `Union' combination from \cite{Komatsu:2008hk}, based on results from WMAP, baryonic acoustic oscillations and high-redshift supernovae; i.e.\ $\Omega_{m}=0.27$, $\Omega_{\Lambda}=0.73$, $h=0.7$, $\Omega_{b}=0.044$, $\sigma_8=0.8$, $n_s=0.96$. These parameters are similar to the recent cosmology results of the Planck collaboration \citep{2013arXiv1303.5076P}, where, considering a combination of data from \emph{Planck}, WMAP, and LSS surveys (showing baryon acoustic oscillations) the parameters were calculated to be: $\Omega_{m} = 0.307\pm0.0042$, $\Omega_\Lambda = 0.692\pm0.010$, $h=0.678\pm0.0077$, $\Omega_{b} = 0.048\pm0.00052$, $\sigma_8 = 0.826\pm0.012$ and $n_s = 0.9608\pm0.00024$. The power spectrum and transfer function used for setting initial conditions was generated using CAMB \citep{Lewis:1999bs}. The {\small CUBEP$^3$M} code's initial condition generator uses first-order Lagrangian perturbation theory (1LPT), i.e.\ the Zel'dovich approximation \citep{1970A&A.....5...84Z}, to place particles in their starting positions. The initial redshift when this step takes place was $z=100$. For a more detailed commentary on the choice of starting redshift for this simulation see \cite{2012arXiv1212.0095W}.

The data handling requirements for analysing the Jubilee simulation were particularly challenging. For each output slice the simulation's particle data totalled around 4TB. These outputs were then analysed and converted into density and then potential fields, as outlined in \S~\ref{sect:meth_isw}, below. The mesh used for the potential fields was 6000$^3$ in size ($(1~h^{-1}$Mpc)$^3$ per cell) so each output slice in redshift for the potentials was 800Gb in size. Overall, the data for the potential fields used in this analysis totalled over 15TB and was reduced from particle data that was 100TB in size. For the weak lensing outputs, discussed in \S~\ref{sect:lensing} below, five derivatives of the potential were calculated, resulting in another 75TB of data.

\section{Methodology}
\label{sect:meth}
\subsection{The ISW effect in the Jubilee simulation}
\label{sect:meth_isw}
The ISW maps are produced adopting a semi-linear approach where the potential is computed exactly in the entire simulation box but its time derivative is computed using linear theory. In a recent work \citet{2010MNRAS.407..201C} demonstrated that this approximation (hereinafter referred to as the LAV approximation, following the terminology of \citeauthor{2010MNRAS.407..201C}) is sufficient to study the ISW on the largest scales with indistinguishable results up to $\ell=40$ in contrast to the exact (non-linear and computationally more expensive) calculation. At $\ell=100$, the LAV approximation under-predicts the real power by nearly an order of magnitude since the LAV does not account for the peculiar velocities that become important at small scales. Nevertheless, most of the ISW effect is concentrated on the largest scales ($\ell<50$) for which the LAV is accurate to within a few percent \citep{2010MNRAS.407..201C}, and the maximum cross-correlation signal is expected to occur around $\ell \sim 10$ for an LSS galaxy survey \citep{2002PhRvD..65j3510C}.

The temperature fluctuations in the CMB induced by the ISW effect can be written as \citep{1967ApJ...147...73S}:

\ba
\label{delta_T_2}
\frac{\Delta T}{T} = \frac{2}{c^2} \int \dot{\Phi}(\mathbf{x},t) dt,
\ea

\noindent where $\dot{\Phi}$ is the derivative of the gravitational potential with respect to time. The potential can be calculated from fluctuations in the density field of the universe via the cosmological Poisson equation:

\ba 
\label{poisson_eqn}
\nabla^2\Phi(\mathbf{x},t)=4\pi G\rho_m(t)a^2\!(t)\delta(\mathbf{x},t) ,
\ea

\noindent where $\rho_m$ is the background matter density and $\delta$ is the `overdensity', defined as:

\ba
\delta(\mathbf{x},t) = \frac{\rho(\mathbf{x},t)-\rho_m(t)}{\rho_m(t)}.
\ea

\noindent In Fourier space equation~\ref{poisson_eqn} is

\ba
\label{phi_eqn}
 -k^2\Phi(\mathbf{k},t) = 4\pi G\rho_m(t)a^2(t)\delta(\mathbf{k},t).
\ea

\noindent Using the present day matter density parameter, $\Omega_{m0} = 8 \pi G \rho_{m0} / 3H_0^2$, and the fact that $\rho_{m}(t) = \rho_{m0}a(t)^{-3}$ we have

\ba
\label{phi_omega_eqn}
\Phi(\mathbf{k};t) = -\frac{3}{2}\Omega_{\mathrm{m}0} \frac{H_0^2}{k^2} \frac{\delta(\mathbf{k};t)}{a(t)}.
\ea

\noindent Differentiating this with respect to time then gives

\ba
\label{phidot_eqn_a}
\dot{\Phi}(\mathbf{k};t) =\frac{3}{2}\Omega_{\mathrm{m}0} \frac{H_0^2}{k^2} \left[\frac{H(t)}{a(t)}\delta(\mathbf{k};t)-\frac{\dot{\delta}(\mathbf{k};t)}{a(t)}\right].
\ea

\noindent For the construction of the ISW maps we make the approximation that the evolution of the overdensity field with time is given by linear theory, where

\ba
\dot \delta(\mathbf{k};t) = \dot D(t)~\delta(\mathbf{k};t=0).
\ea

\noindent and $D(t)$ is the growth factor \citep{1977MNRAS.179..351H}. We can substitute for $\dot{\delta}(\mathbf{k};t)$ in equation~\ref{phidot_eqn_a} resulting in

\ba
\label{phidot_eqn_b}
\dot{\Phi}(\mathbf{k};t) = \frac{3}{2}\Omega_{m0}\frac{H_0^2}{k^2}\frac{\delta(\mathbf{k},t)}{a(t)} H(t)\left(1-\beta(t)\right),
\ea

\noindent where $\beta(t) = d\mathrm{ln}D(t)/d\mathrm{ln}a(t)$. Finally, combining equations~\ref{phi_omega_eqn} and~\ref{phidot_eqn_b} results in

\ba
\label{phidot_phi}
\dot{\Phi} = -\Phi H(t) [1-\beta(t)],
\ea

\noindent which is valid in both real and Fourier space. 

To calculate the ISW effect in the Jubilee simulation, we first produce a smoothed overdensity field, $\delta(\mathbf{x},t)$, from the particle outputs from 20 timeslices between $z=0$ to $1.4$. The overdensity field is calculated using a Cloud-In-Cell (CIC) smoothing kernel \citep[see, for example,][]{Hockney:1988:CSU}. Then, from the $\delta(\mathbf{x},t)$ field we use the Multiple Fourier Transform (MFT) method \citep{Hockney:1988:CSU} to calculate the potential field $\Phi(\mathbf{k},t)$. This follows the steps outlined above, solving the Poisson equation in the Fourier domain. We then produce maps of the real-space potential in redshift shells given by the distribution of the simulation time slices, which totalled 20 between $z = 0$ to $1.4$. To produce the maps, we traced rays from a centrally-located observer through each of the cells of the potential field. The potential for each shell, integrated along lines of sight in this manner, was then projected onto the sky using {\small HEALPix}\footnote{\url{http://healpix.jpl.nasa.gov}} \citep{2005ApJ...622..759G}. We applied a linear interpolation between the different slices in order to account for potential values at intermediate redshifts (the net effect of interpolating versus not interpolating is $<1\%$ on the final results). From these outputs we then used equation~\ref{phidot_phi} to calculate $\dot{\Phi}$ and calculated the ISW effect using equation~\ref{delta_T_2}.

\subsection{LRG catalogue construction}

For correlating the ISW with LSS, we first need to create a suitable catalogue of tracers of the dark matter density field. For the ISW-LSS signal, as we shall see, a population of tracers that exist between redshifts of $z\sim0.1$ to $1.0$ create the strongest signal. LRGs are, therefore, very useful because they are detectable across the range in question due to their high luminosities. The majority of LRGs reside in haloes that have masses in excess of 10$^{13}~h^{-1}\mathrm{M}_{\odot}$ \citep{2009ApJ...707..554Z}. They are typically the Brightest Cluster Galaxy (BCG) in their cluster and are located at the centre of their parent dark matter haloes \citep{2009ApJ...707..554Z,2012ApJS..199...34W,2012MNRAS.426.2944Z} (although note that the corollary is not true: BCGs are not typically LRGs: \cite{2012ApJS..199...34W} show that $\sim25\%$ of BCGs are LRGs). Complications with this scenario arise in high mass clusters where there exists a fraction of LRGs ($\sim 5\%$) that are satellites \citep{2009ApJ...707..554Z}. In this study we ignore satellite LRGs and model only a population of central LRGs in our dark matter haloes.

\subsubsection{Halo finding}

To create an LRG catalogue we need to find dark matter haloes in our simulation. We used {\small CUBEP$^3$M}'s own on-the-fly SO halofinder (hereafter `CPMSO') to do this. This halo finder is based on the Spherical Overdensity (SO) algorithm \citep{1994MNRAS.271..676L} and the full details of how the finder works can be found in \cite{HarnoisDeraps:2012he}. A comparison of the mass function results from the CPMSO halofinder to the Amiga Halofinder (AHF) \citep{Gill:2004km,Knollmann:2009pb}, can be found in \cite{2012arXiv1212.0095W}. Results from the CPMSO and AHF halofinders and from a Friends-of-Friends halofinder specifically applied to the Jubilee simulation can be found in \cite{2013arXiv1305.1976W}. As the CPMSO halofinder runs on-the-fly within the $N$-body code we can relatively easily output data for haloes across a number of redshifts. These were chosen to match the output redshifts for our potential fields.

\subsubsection{Modelling of central LRGs in haloes}

We resolve galaxy size haloes in the Jubilee simulation down to $\sim10^{12}~h^{-1}\mathrm{M}_\odot$ but not all haloes of this mass and above contain LRGs. To model a population of LRGs from our dark matter haloes we applied part of a Halo Occupation Distribution (HOD) model to select which haloes host LRGs. The model we used was that of \cite{2009ApJ...707..554Z} who studied a sample of LRGs from the Sloan Digital Sky Survey \citep{2005ApJ...633..560E} from $z=0.16$ to $0.44$. We apply, specifically, the prescription laid out in appendix B of \cite{2009ApJ...707..554Z} which gives the average occupation function \citep[based on ][]{2005ApJ...633..791Z} for \textit{central} LRGs as

\ba
\label{n_occ:eqn}
\langle \mathrm{N_{cen}} \rangle_\mathrm{M} = \frac{1}{2}\left [1+\mathrm{erf}\left(\frac{\mathrm{log~M}-\mathrm{log~M_{min}}}{\sigma_{\mathrm{log~M}}}\right)\right],
\ea

\noindent where erf is the error function, $\langle \mathrm{N_{cen}} \rangle_\mathrm{M}$ is the average number of central LRGs in a halo of mass M, $\sigma_{\mathrm{log ~M}}$ controls the width in the log~M-N relation and $\mathrm{M_{min}}$ is a characteristic minimum mass of hosts with central galaxies. The central LRGs follow a nearest integer probability distribution. This model allows us to populate central LRGs in our haloes using a random number generator.

\begin{figure}
  \begin{center}
    \includegraphics[width=3.2in]{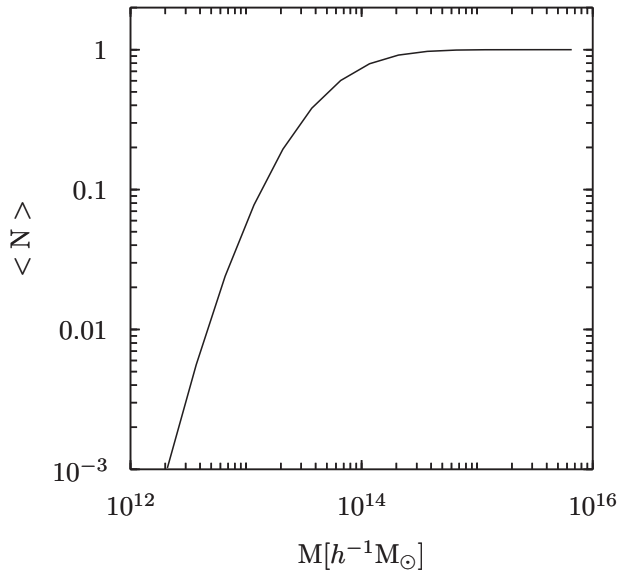}
  \end{center}
  \caption{The average occupation number of central LRGs in host haloes of mass $M$, based on the model of \citet{2009ApJ...707..554Z}.
    \label{HOD:fig}}
\end{figure}

The variables in equation~\ref{n_occ:eqn} were calculated by \cite{2009ApJ...707..554Z}, based on a volume-limited sample of LRGs with a redshift range of $z=0.16$ to $0.44$. The absolute magnitude cut off for this sample was based on a rest frame \textit{g}-band magnitude of $M_g < -21.2$ (note that we refer to masses as unitalicised, M, and magnitudes as italicised, $M$) which was calculated at $z=0.3$ for all LRGs and included corrections for evolution. The HOD parameters were found to be: $\mathrm{log~M_{min}}=13.673 \pm 0.06~h^{-1}\mathrm{M_{\odot}}$ and $\sigma_{\mathrm{logM}} = 0.621 \pm 0.07~h^{-1}\mathrm{M_{\odot}}$. \noindent In populating our haloes with LRGs we make the additional assumption that the above error bars in the model parameters -- which are given to $1\sigma$ -- can be modelled using a Gaussian distribution, which we use to introduce a similar error into our catalogue so as to mimic this uncertainty in the model. The halo occupation function for our haloes is shown in Figure~\ref{HOD:fig}. As can be seen from this plot there is a sharp drop-off in halo occupation below $10^{14}~h^{-1}\mathrm{M}_{\odot}$, to the extent that $10^{13}~h^{-1}\mathrm{M}_{\odot}$ haloes contain, on average, 0.05 LRGs.

\subsubsection{Luminosity modelling}

Now we have a population of LRGs in our haloes we need to assign properties to them, most importantly their luminosities. To do this, we rely solely on the mass of the host haloes. The results presented in \cite{2009ApJ...707..554Z} indicate that the entire population of LRGs in their sample obeys the simple relation $L \propto \mathrm{M}^{0.66}$. Unfortunately, this is an inadequate prescription for assigning luminosities to our LRGs as, over the entire mass range of our host haloes, it results in too many unrealistically bright LRGs. A more detailed description of the \textit{L}-M relationship is shown in Figure 3 of \cite{2009ApJ...707..554Z}, which implies that at higher host halo masses the luminosity of LRGs does not scale as steeply as for lower masses. \cite{2009ApJ...707..554Z} discuss this result and make comparisons to other work which shows a similar trend. For our modelling we adopt, based on their figure, a relationship between mass and luminosity of the form:

\ba
\label{eqn:l_m_alpha}
L \propto \mathrm{M}^\alpha,
\ea

\noindent where the parameter $\alpha$ is given by

\ba
\label{eqn:alpha}
    \alpha =
\begin{cases}
    1& \text{if } \mathrm{M}\leq 5\times 10^{11}~h^{-1}\mathrm{M}_{\odot}\\
    0.5& \text{if } 5\times 10^{11}~h^{-1}\mathrm{M}_{\odot} \leq \mathrm{M} < 5\times 10^{12}~h^{-1}\mathrm{M}_{\odot}~.\\
    0.3& \text{if } \mathrm{M}\geq 5\times 10^{12}~h^{-1}\mathrm{M}_{\odot}
\end{cases}
\ea

\begin{figure}
  \begin{center}
    \includegraphics[width=3.2in]{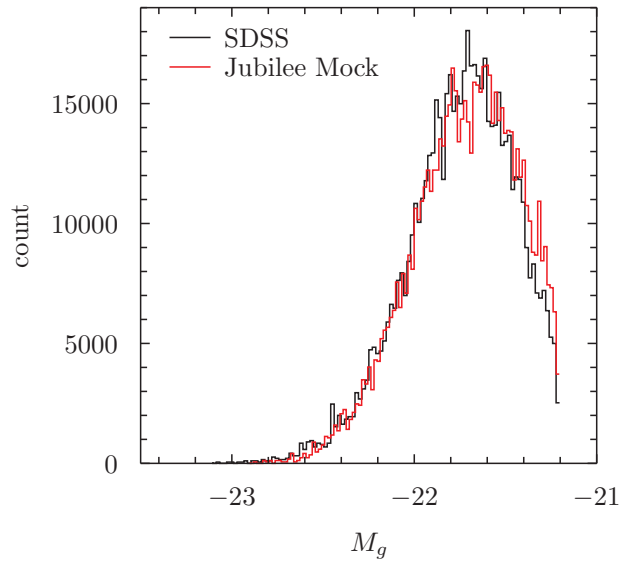}
  \end{center}
  \caption{Histogram comparing SDSS LRGs with Jubilee mock LRGs. The full dataset of SDSS DR7 LRGs from \citet{2010ApJ...710.1444K} is plotted together with a random subsample of Jubilee Mock LRGs with the same total number count. The SDSS data is taken from a redshift range of $z=0.16$ to $0.44$ with a g-band absolute magnitude range of $M_g < -21.2$ (calibrated at $z=0.3$). Jubilee mock data is taken from the $z=0.3$ output slice. 
    \label{Mg_SDSS:fig}}
\end{figure}

\noindent This, combined with the comoving number density of LRGs in the sample, allows luminosities to be allocated to our LRGs in a manner that produces correctly the observed luminosity distribution of SDSS LRGs. We show a comparison of our model to the $M_g < 21.2$ SDSS sample in Figure~\ref{Mg_SDSS:fig}. The SDSS data was based on the catalogue of \cite{2010ApJ...710.1444K}, who closely match the previous catalogue of \cite{2005ApJ...633..560E}.

We apply this model to our data past the $z = 0.44$ limit of the modelling dataset. This is in order to create a base set of LRGs from which to work with from all redshift slices in the simulation. For LRGs that exist at higher redshifts, this base dataset may require corrections. For example, the various details of specific pipelines from observational catalogues can be readily incorporated onto this data, including any offsets from the LRG catalogue modelled here.

\subsubsection{Other LRG properties}

The halo catalogue contains information on the locations dark matter density peaks. The question of whether this corresponds to the locations of cluster BCGs has been recently studied by \cite{2012MNRAS.426.2944Z}, who used strong lensing to probe the underlying dark matter distributions in 10,000 SDSS clusters. Their results show a small offset, with no preferred orientation, to the locations of BCGs from the dark matter density peaks. We apply their results to our dark matter halo catalogues in order to introduce this discrepancy between central LRGs, which we assume to be the BCGs in their particular haloes, and the underlying matter field that seeds the gravitational potential.

\begin{figure}
  \begin{center}
    \includegraphics[width=3.2in]{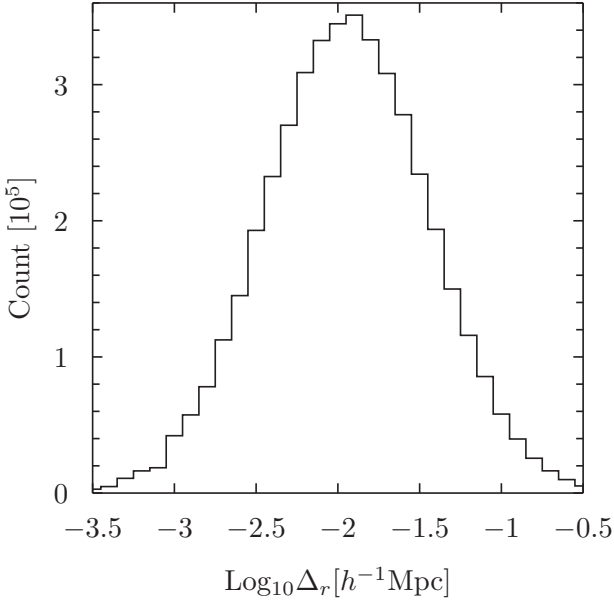}
  \end{center}
  \caption{Distribution of random offsets between halo centres and the LRG locations, based on the results of \citet{2012MNRAS.426.2944Z}. The data shown here are based on 438,000 LRGs at $z = 0.5$.
    \label{zit:fig}}
\end{figure}

The results of \cite{2012MNRAS.426.2944Z} showed that the scatter between the BCG location and density peaks are distributed log-normally in random directions via: $\mathrm{log}_{10}\Delta_r=-1.895_{-0.004}^{+0.003}~h^{-1}\mathrm{Mpc}$. We produced a random scatter based on this and show the effect using a histogram in Figure~\ref{zit:fig}. This figure should be directly compared to Figure 5 from \cite{2012MNRAS.426.2944Z}. We note that they observed a potential trend with redshift to this scatter, that is, that the peak in Figure~\ref{zit:fig} would sit at $\sim-2.5$ for haloes at $z\sim0.15$ and would evolve to $\sim-1.7$ for haloes at $z\sim0.6$. However, the error in these results is large, being $\sim\pm0.5$, and the trend of the evolution appears to be flattening out towards higher redshifts. Considering these facts we do not attempt to parameterise the offset with redshift.

Finally, the bulk velocities of the haloes are taken to be the same as the LRG velocities. This is an assumption and one that is likely to be incorrect to a certain degree (as illustrated partially by \cite{2013ApJ...762..109B} who utilise a phase-space halofinder to illustrate that halo cores frequently have an offset in velocity relative to the bulks of the parent haloes). In constructing sky maps of the LRGs with magnitude cuts imposed it is possible to consider either a redshift that has been shifted due to the peculiar velocity of the LRGs or one that has not. In this paper we consider LRGs with a Doppler-included redshift.

\subsubsection{Simulated sky catalogues}

With a complete set of LRGs in each of our output redshift slices it is possible to impose cuts to the catalogue in an attempt to mimic different observational catalogues. As an example we model here the SDSS sample of \cite{2005ApJ...633..560E}, which is a natural choice because this is the sample on which \cite{2009ApJ...707..554Z} based their modelling. To create this catalogue we simply apply the magnitude cut from \cite{2005ApJ...633..560E} onto our data. We assume that the catalogue covers the full sky and refer to it from this point as the `SDSS mock' catalogue. In principle, other catalogues can be simulated by adopting constraints on magnitude and sky coverages, combined with subtleties such as completeness and scatter in photometric redshifts etc.

\begin{figure}
  \begin{center}
    \includegraphics[width=3.2in]{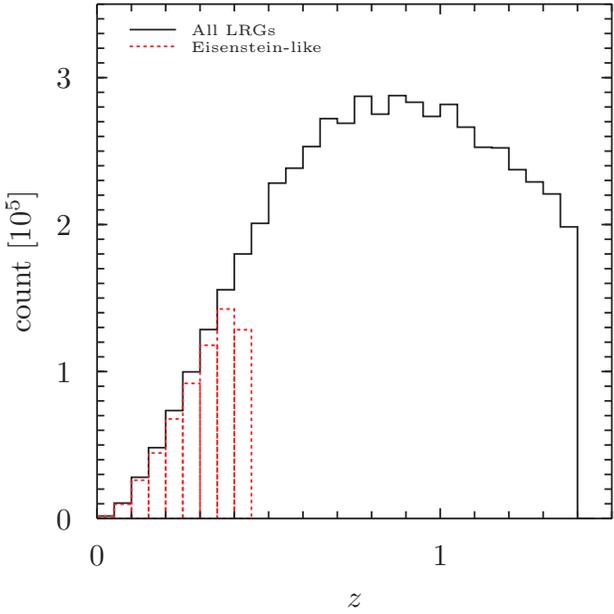}
  \end{center}
  \caption{Histogram of LRG number counts. The count for our entire mock catalogue is shown in black. In red (dashed) we show the counts in our SDSS mock catalogue, which approximates the properties of the sample of \citet{2005ApJ...633..560E} in the full sky.}
    \label{lrg_hist:fig}
\end{figure}

We show a histogram of LRG counts for both the SDSS mock catalogue and our entire sky catalogue of LRGs in Figure~\ref{lrg_hist:fig}. The drop-off in LRG counts at low redshifts is due to the smaller volumes being sampled. The drop-off for $z>1$ occurs because of the LRGs becoming rarer as the halo mass function evolves, cutting down the number of appropriately massive hosts as it does so.

\subsection{Weak lensing maps}
\label{sect:lensing}

The weak lensing potential is proportional to the ISW potential. As such we can calculate both the weak lensing and ISW effects from the same data. The lens equation is given by

\ba  
\vect{\beta} = \thetavec - \vect{\alpha}(\thetavec,m(\thetavec))  
\label{eq_lens}  
\ea

\noindent where $\vect{\alpha}(\thetavec)$ is the deflection angle created by the lens  which depends on the observed positions, $\thetavec$. We can write a dimensionless, integral version of equation~\ref{poisson_eqn} as

\ba
\Phi(r_p) = -\frac{G}{c^2} \int \frac{\rho(r_p-r'_p)}{|r_p-r'_p|}d^3r'_p,
\label{eq_Phi1}
\ea

\noindent where $r_p=(x,y,z)$ is a position in the simulation box in physical units. Now we can define a new scalar (and adimensional) lensing potential in a given direction $\theta$:

\ba
\psi(\theta) = \frac{2D_{ls}}{D_lD_s} \int \Phi(D_l\theta,z)dz,
\label{eq_psi}
\ea

\noindent where $r_p^2 = (D_l\theta)^2 + z^2$. The distances $D_{ls}$, $D_l$, and $D_s$ are the angular distances from the lens to the source, the distance from the observer to the lens and the distance from the observer to the source respectively. The relevant lensing quantities we are interested in are then obtained from the derivatives of $\psi$. The derivatives are made with respect to the components of $\theta$, i.e.\ ($\theta_1,\theta_2)$. The deflection angle $\vect{\alpha}=(\alpha_1,\alpha_2)$ is given by the divergence (or first derivatives) of $\psi$ and both the shear, $\vect{\gamma}=(\gamma_1,\gamma_2)$, and convergence, $\kappa$, are defined in terms of the second partial derivatives:

\ba
\alpha_1(\thetavec) 	&=& \psi_{1}, \\
\alpha_2(\thetavec) 	&=& \psi_{2},\\
\gamma_1(\thetavec) 	&=& \frac{1}{2}(\psi_{11} - \psi_{22})
                        = \gamma(\thetavec)\cos[2\varphi], \\
\gamma_2(\thetavec) 	&=& \psi_{12} = \psi_{21} =\gamma(\thetavec)\sin[2\varphi],
\ea

\noindent where $\gamma(\thetavec)$ is the amplitude of the shear and $\varphi$ its orientation and

\ba
\psi_{i} = \frac{\partial\psi}{\partial\theta_i},
\ea
\ba
\psi_{ij} = \frac{\partial^2\psi}{\partial\theta_i\partial\theta_j}.
\ea

\noindent The amplitude and orientation of the shear are given by
\ba
\gamma = \sqrt{\gamma_1^2 + \gamma_2^2},
\ea
\ba
\varphi = \frac{1}{2} \mathrm{atan}\left(\frac{\gamma_2}{\gamma_1} \right).
\ea

\noindent The convergence is

\ba
\kappa(\thetavec) = \frac{1}{2}(\psi_{11} + \psi_{22}).
\label{eq_kappa1}
\ea

\noindent Finally the magnification, $\mu$, is
\ba
\mu = \frac{1}{(1-\kappa)^2 - \gamma^2}.
\ea

\noindent All these relevant quantities that describe the lensing effect can be obtained by combining the 5 derivatives, $\psi_1$, $\psi_2$, $\psi_{11}$, $\psi_{22}$, and $\psi_{12}=\psi_{21}$. For our particular case, given that we have simulation data in 3-dimensional Cartesian coordinates, it is convenient to express the derivatives of the lensing potential (originally with respect to the angle $\theta=(\theta_1,\theta_2)$) with respect to the physical coordinate $r = (x,y) = \theta D_l$. In these coordinates $\nabla_\theta = D_l \nabla_r$. The first and second derivatives with respect to $\theta$ of equation (\ref{eq_psi}) can be rewritten in terms of derivatives with respect to $r=(x,y)$ as
\ba
\vect{\nabla}_\theta \psi(\theta) = F_{l1} \int \vect{\nabla_r}\Phi(x,y,z)dz,
\label{eq_psi_r1}
\ea
\ba
\nabla^2_\theta \psi(\theta) = F_{l2} \int \nabla^2_r \Phi(x,y,z)dz,
\label{eq_psi_r2}
\ea
\noindent where $F_{l1}=2D_{ls}/D_s$ and $F_{l2}=2D_lD_{ls}/D_s$. From our simulation outputs we convert the various Cartesian datasets into their sky projections by adopting the following coordinate system:

\ba
x &=& \sin(\theta)\sin(\phi)\nonumber, \\
y &=& \sin(\theta)\cos(\phi), \\
z &=& \cos(\theta)\nonumber,
\ea

\noindent where $\theta_1 = \theta$ and $\theta_2 = \phi$. We then compute the derivatives of $\Phi$ in the $(\theta_1,\theta_2)$ coordinate system. We assume in our analysis that the source object behind the lens is at a redshift of $z=10$. All of our maps can be easily rescaled to simulate source objects that are at any redshift behind our lensing density fields, for example the CMB.

\subsection{Online databases}

All our LRG data will be publicly available online at \url{http://jubilee-project.org}. An SQL database has been set up so that the data can be queried to suit the requirements of individual users. In addition to LRG catalogues we will also be providing halo catalogues, void catalogues, an NVSS-like radio catalogue, as well as sky maps including lensing maps and density fields.

\section{Results}
\label{sect:res}

\subsection{ISW}

\begin{figure*}
  \begin{center}
    \includegraphics[width=6in]{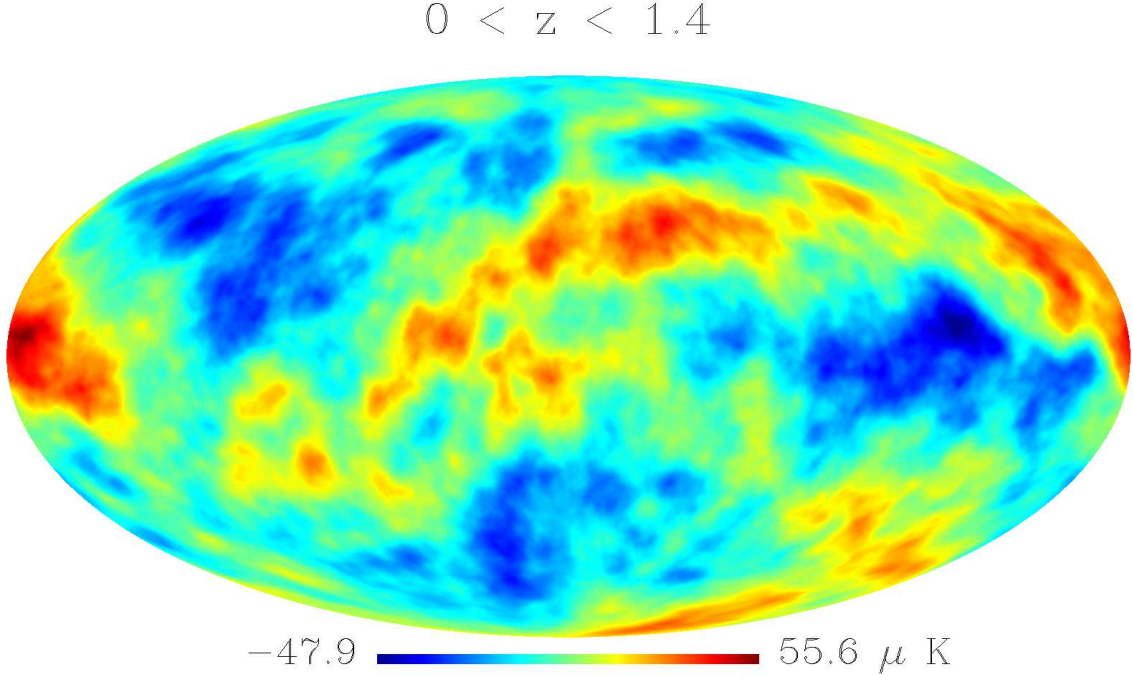}
  \end{center}
  \caption{The full-sky map of the predicted secondary CMB anisotropies due to the ISW effect from structures between redshifts of $z=0$ to $1.4$. The map is obtained by ray-tracing through the simulation potential field using the LAV approximation, as explained in \S~\ref{sect:meth_isw}. The map is shown in Mollweide projection at a resolution of $\mathrm{Nside}=512$. The dipole contribution has been removed.}
    \label{ISW:fig}
\end{figure*}

\begin{figure*}
\begin{center}$
\begin{array}{cc}
\includegraphics[width=3.5in]{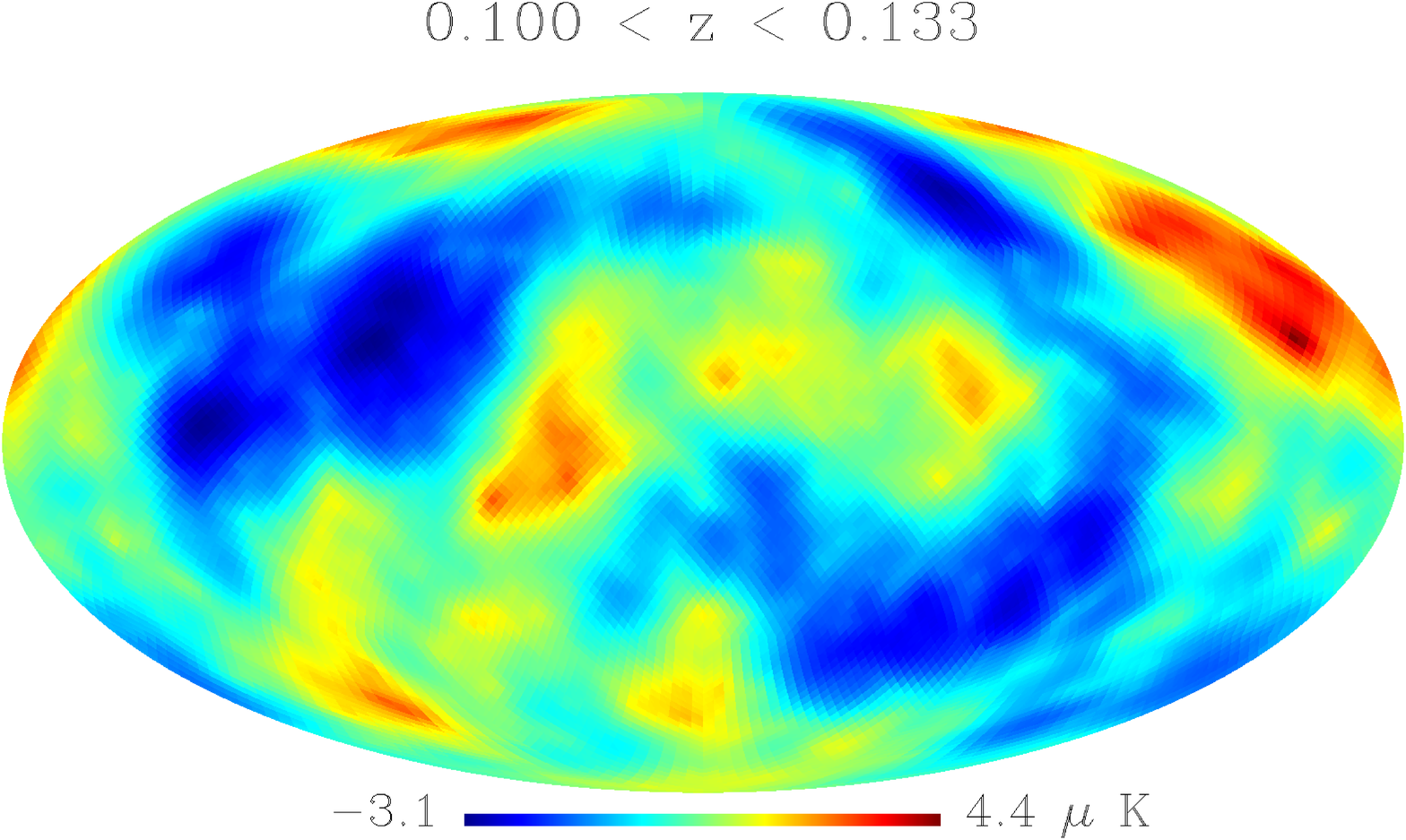}
\includegraphics[width=3.5in]{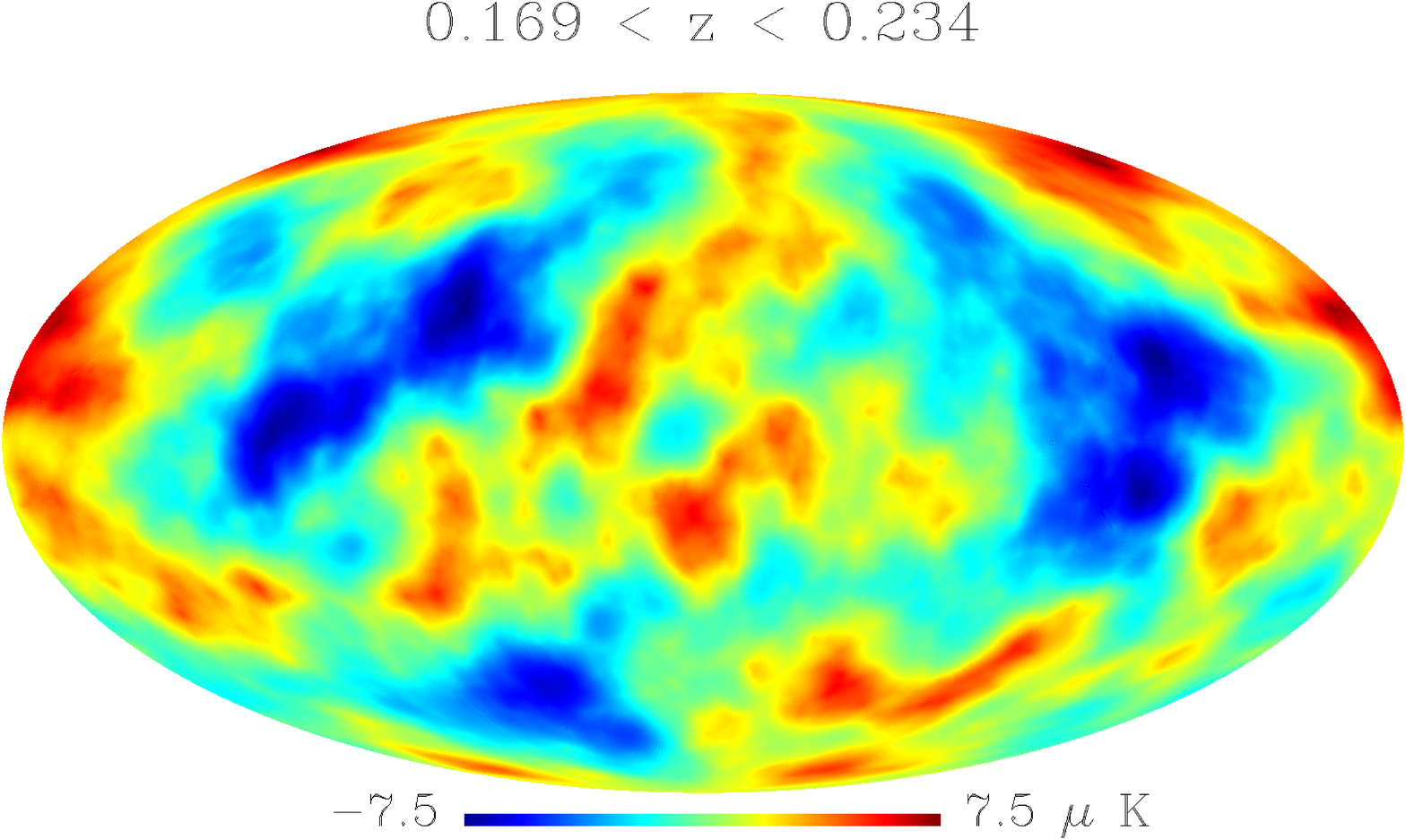} \\
\includegraphics[width=3.5in]{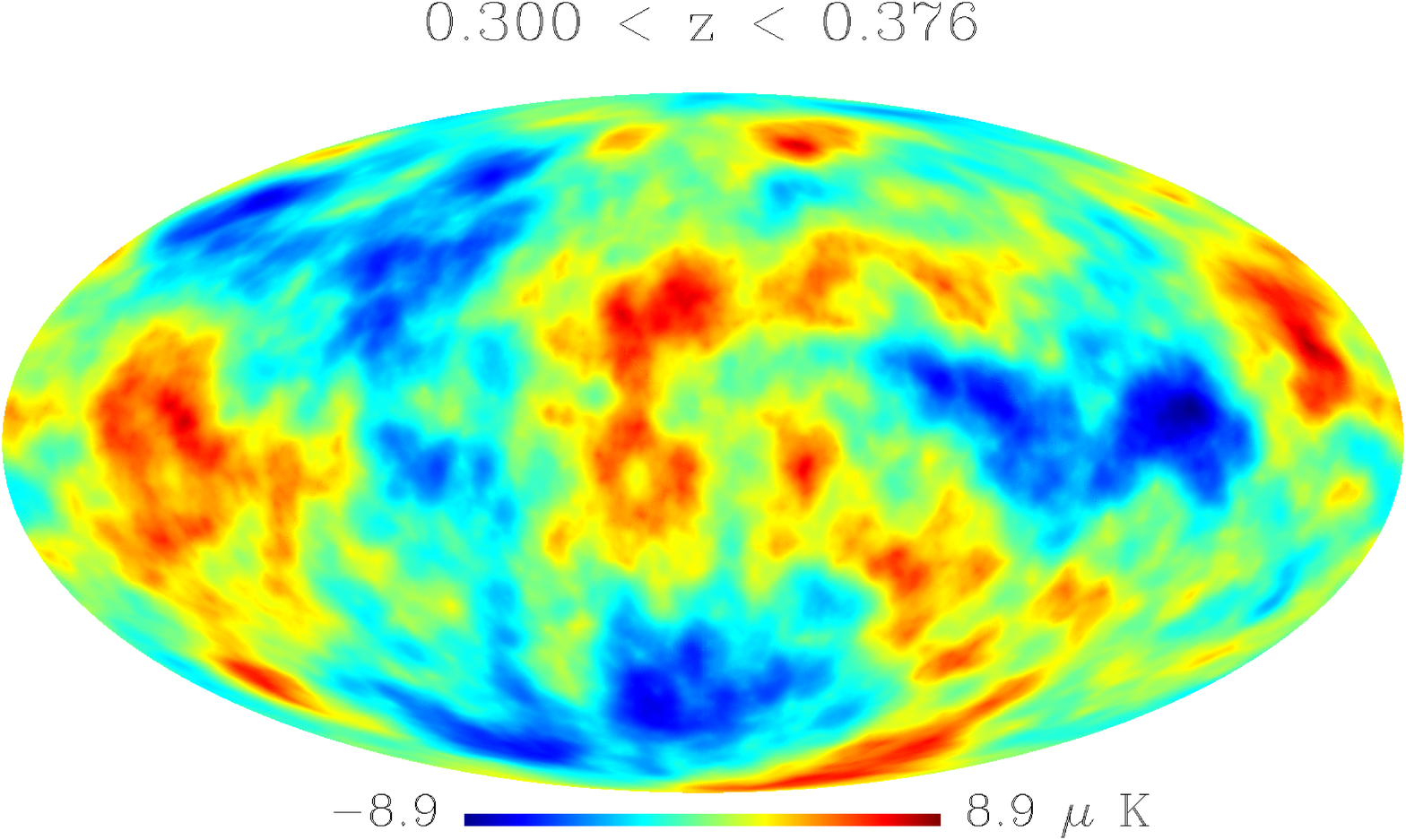} 
\includegraphics[width=3.5in]{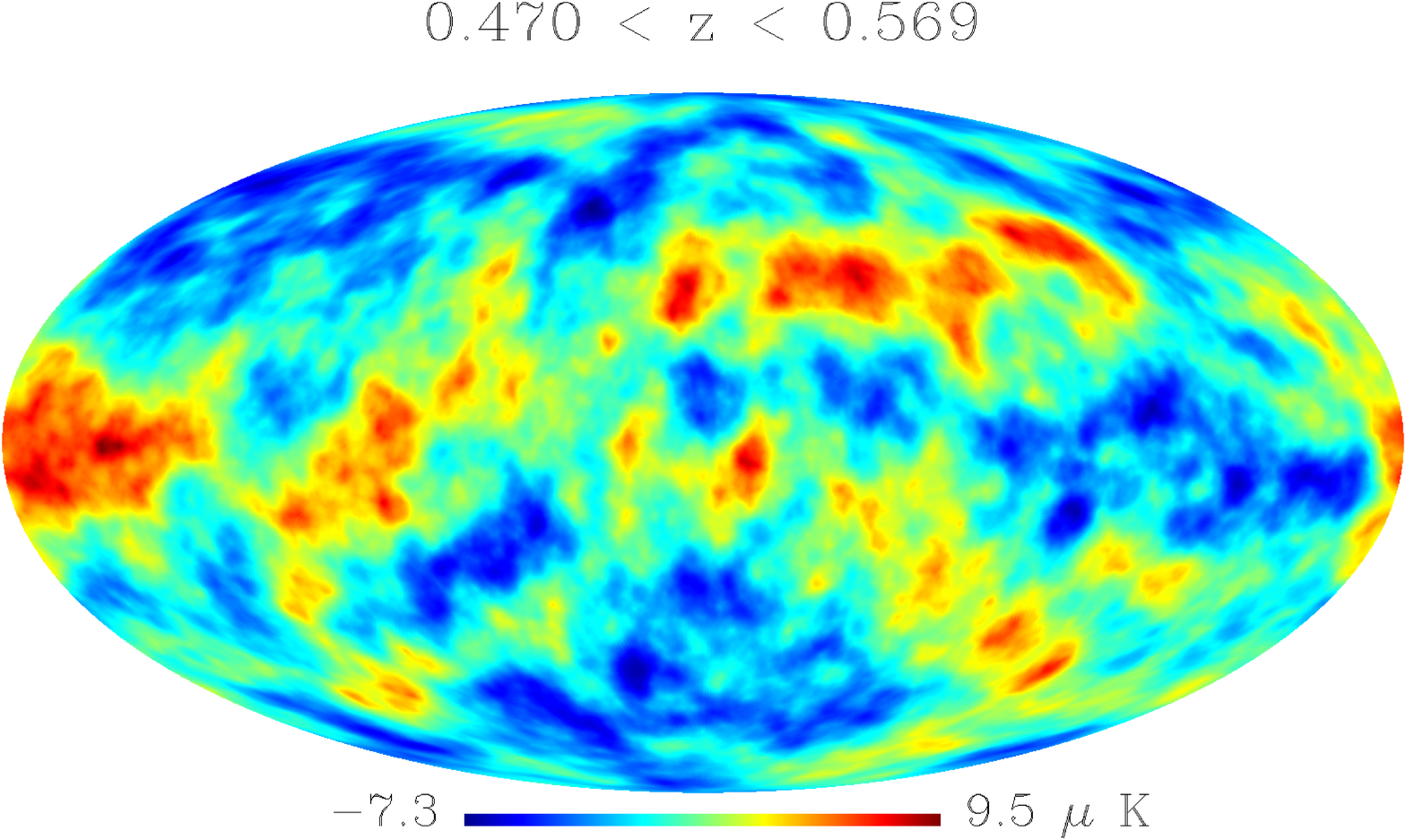} \\
\includegraphics[width=3.5in]{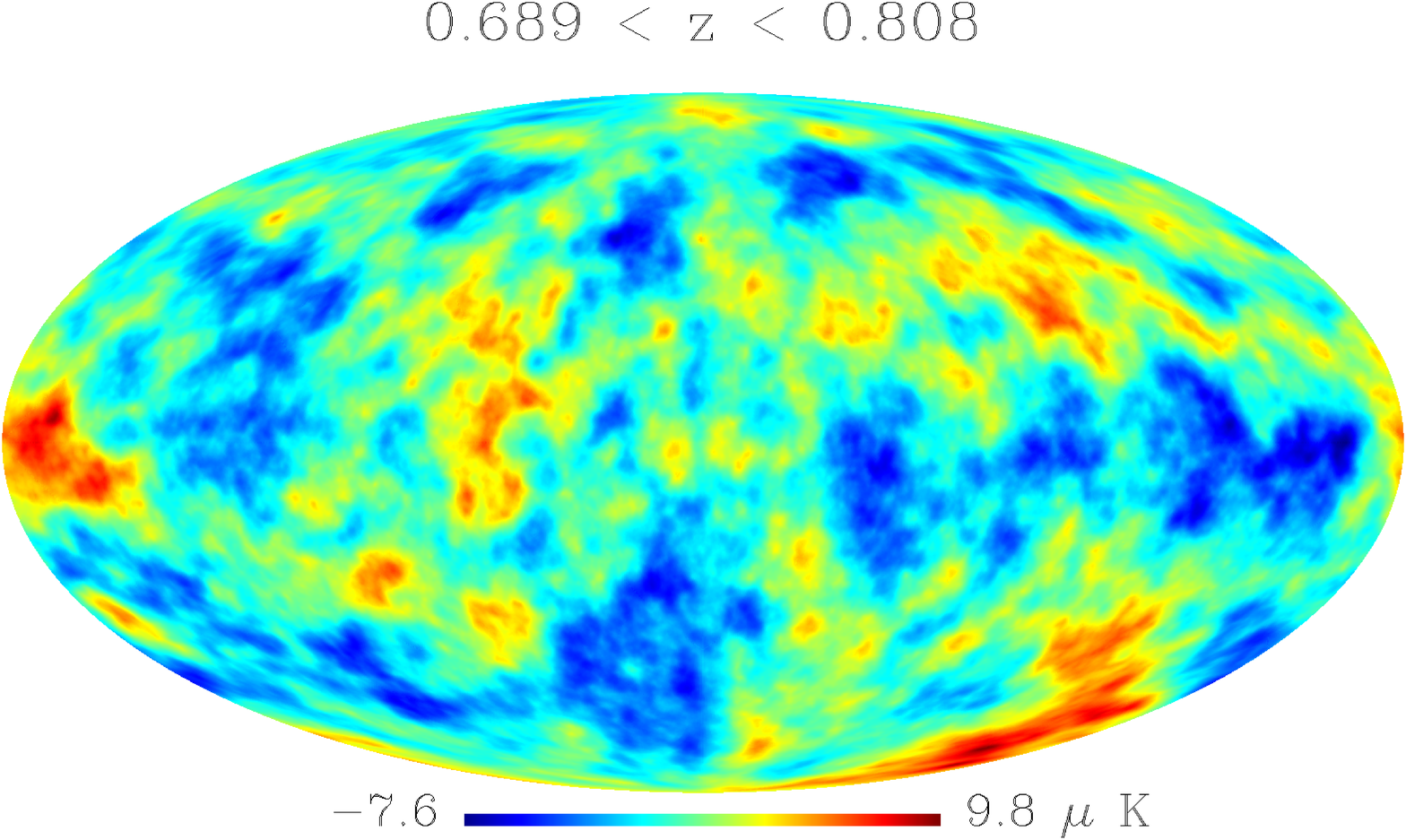} 
\includegraphics[width=3.5in]{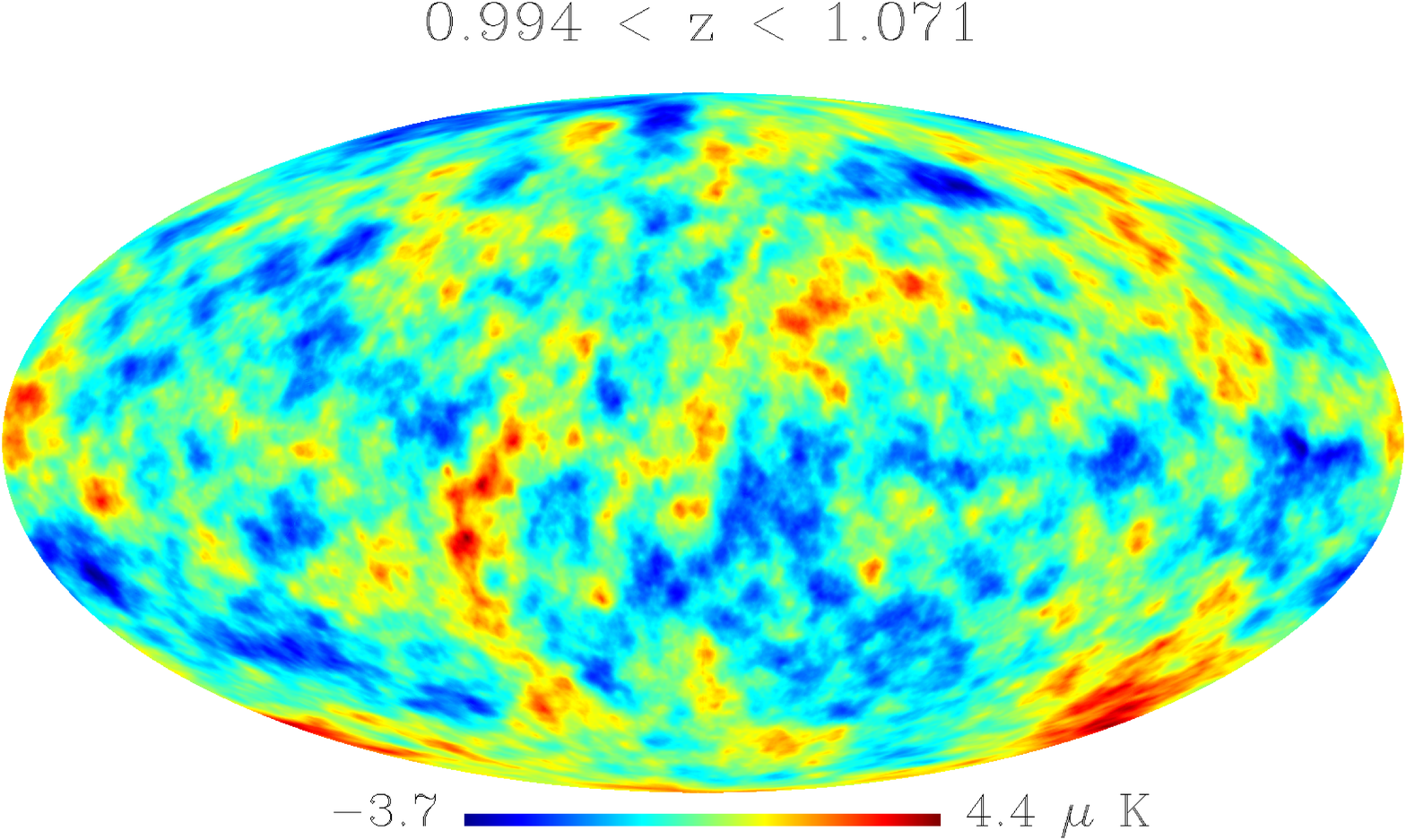} \\
\includegraphics[width=3.5in]{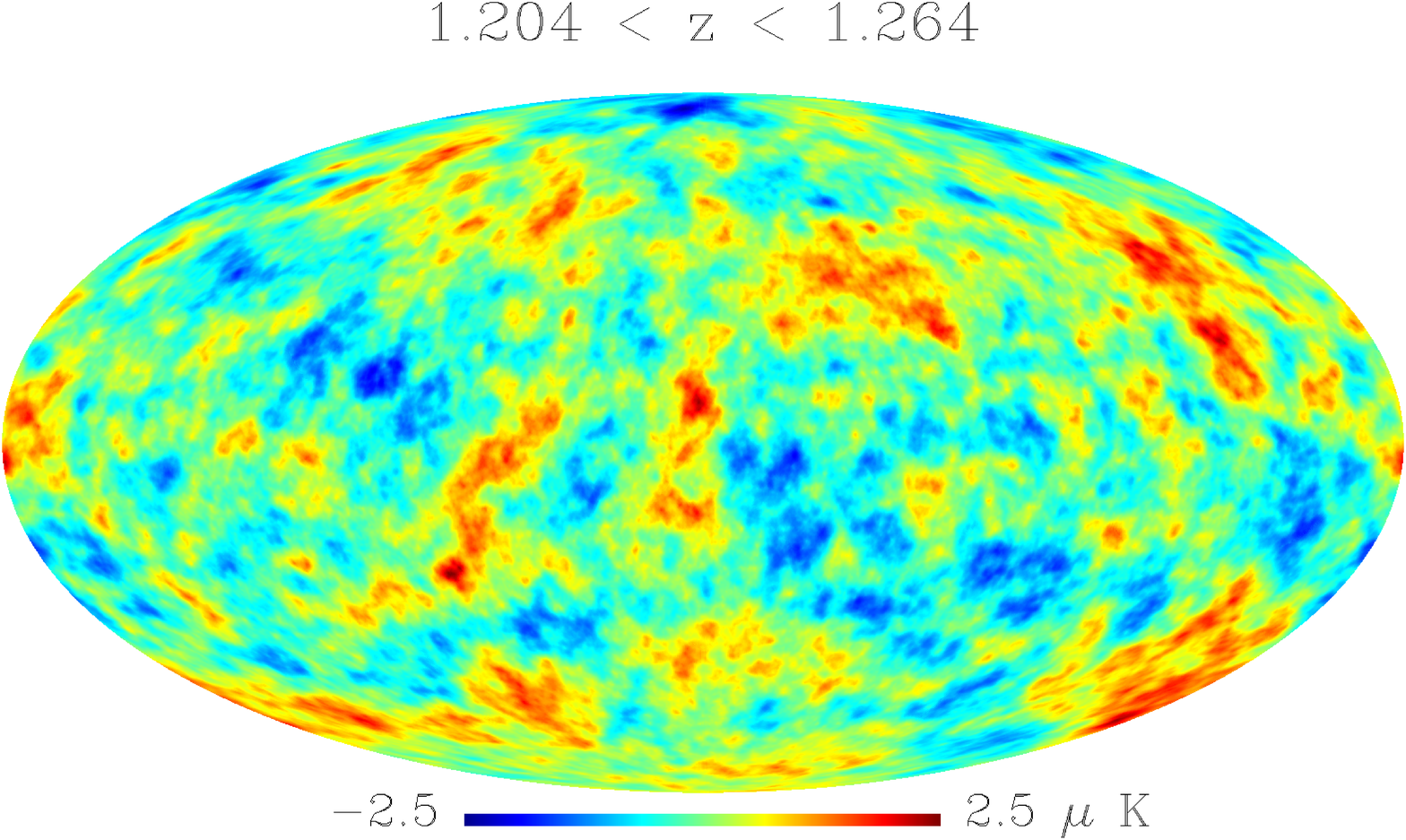} 
\includegraphics[width=3.5in]{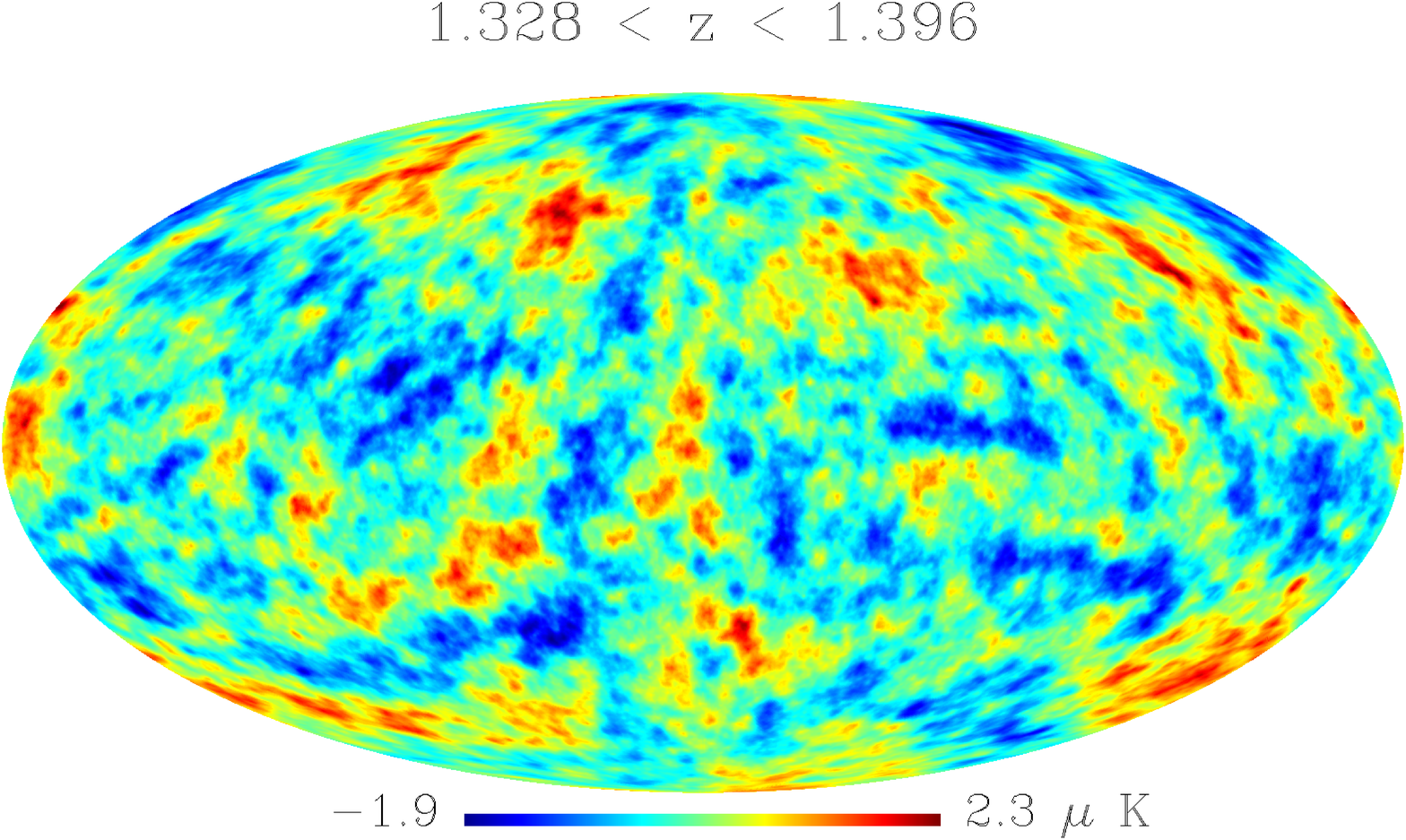} \\
\end{array}$
\end{center}
\caption{Full-sky maps of the predicted secondary CMB anisotropies due to the ISW effect from structures between selected output redshifts. The maps are obtained by ray-tracing through the simulation potential field using the LAV approximation, as explained in \S~\ref{sect:meth_isw}. The maps are shown in Mollweide projection with resolution $\mathrm{Nside} = 32,~128,~256$ and $512$ for redshifts of 0.100-0.133, 0.169-0.234, 0.320-0.569 and 0.689+, respectively. Dipoles have been removed from all maps.}
    \label{ISW_multi:fig}
\end{figure*}

\begin{figure}
  \begin{center}
    \includegraphics[width=3.4in]{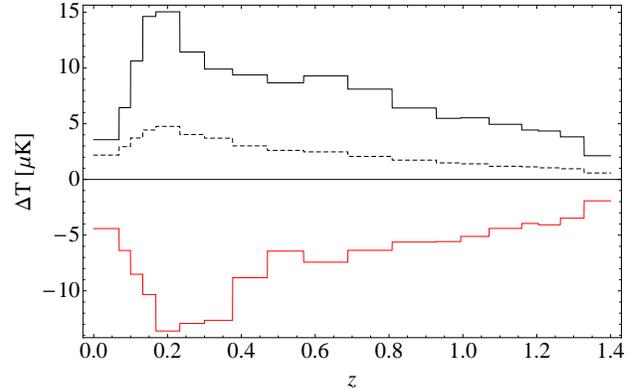}
  \end{center}
  \caption{The black (red) lines show the temperature shift $\Delta \mathrm{T}$ for the hottest (coldest) pixel in each output ISW map for different redshift shells. To account for the different physical lengths of the shells the $\Delta \mathrm{T}$ values have been scaled to indicate the temperature shift that would be produced by a redshift shell of fixed width $\Delta z=0.1$ placed at the central redshift of each bin. The dashed black line shows the $1\sigma$ fluctuation for all pixels in each output map, scaled in the same way.}
    \label{ISW_hist:fig}
\end{figure}

In Figure~\ref{ISW:fig} we show the projected dipole-subtracted ISW all-sky map from redshift $z=0$ to $1.4$. The negative blue regions correspond to projected under-dense regions where the dark-energy-driven acceleration of the expansion results in a net loss of energy for the CMB photons. On the other hand, when the CMB photons cross an over-dense region (red), the decaying potentials result in a net gain of energy for that photon. This map was constructed from a number of redshift shells and we show some of the maps from these shells in Figure~\ref{ISW_multi:fig}. This figure illustrates the varying imprint of the ISW anisotropies across redshifts. Lower redshifts show fluctuations over much larger areas of the sky than at higher redshifts, due to the changing angle that objects subtend in the sky at different redshifts. In addition, the amplitude of the anisotropies varies significantly across the redshift shells as illustrated in Figure~\ref{ISW_hist:fig}, where we show $\Delta \mathrm{T}$ values that have been scaled to indicate the temperature shift that would be produced by a redshift shell of fixed width $\Delta z=0.1$ placed at the central redshift of each bin. We also show the $1\sigma$ fluctuation for all pixels in each output map, scaled in the same way. This plot shows the expected trend that the anisotropies grow stronger as the dark energy component of the cosmological fluid increases in influence. The drop in maximum amplitude and variance of the signal at low redshifts ($z<0.2$) is an effect of sample variance, as the output maps are dominated by a small number of very local structures subtending large angles at the observer location, despite the fact that globally in the simulation $\dot\Phi$ is larger at these times.

\subsubsection{ISW power}
\label{sect:isw_pow}

\begin{figure}
  \begin{center}
    \includegraphics[width=3.2in]{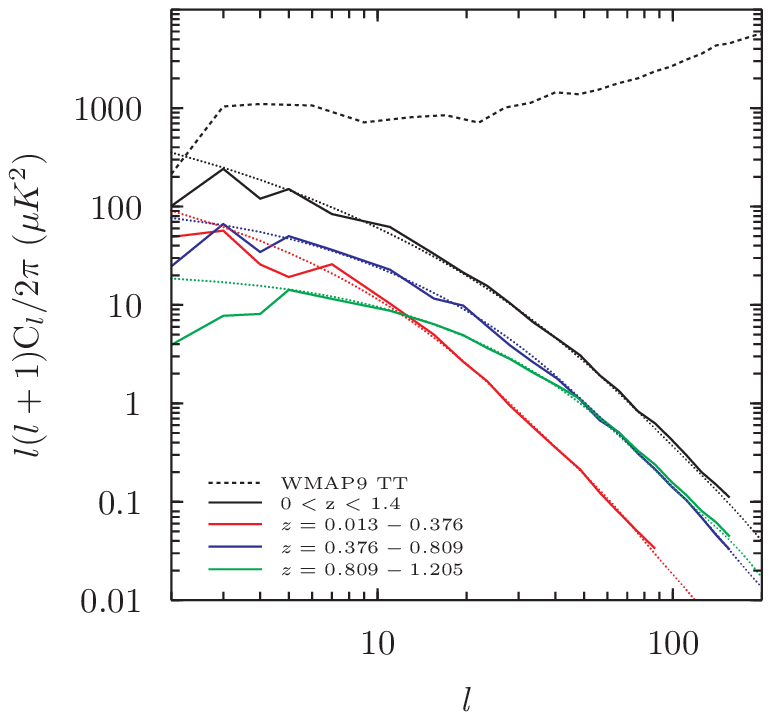}
  \end{center}
  \caption{The power spectrum of the temperature anisotropies that arise from the ISW effect. Power from individual redshift bins are shown along with the full integrated ISW power spectrum (from $z=0$ to $1.4)$ and the 5-year CMB TT power spectrum of \citet{Dunkley:2008ie}. Linear theory predictions are shown as dotted lines.}
    \label{ISW_pow:fig}
\end{figure}

The power spectrum of ISW-induced temperature anisotropies is shown in Figure~\ref{ISW_pow:fig}. The spectrum shows a maximum at low-$\ell$. At higher $\ell$s the slope of the spectrum follows a power law. This is the expected result using the LAV approximation. \cite{2010MNRAS.407..201C} performed a detailed study of the contribution of the velocity field to the ISW effect showing that the LAV power spectrum falls below that of the full ISW effect for higher values of $\ell$, such that the amplitude of the LAV ISW effect at $\ell \sim 100$ is around $50\%$ of the full ISW amplitude, dropping down from $\sim100\%$ at $\ell \le 40$ \citep[see Figure 17 of][]{2010MNRAS.407..201C}. The under-representation of the ISW effect by the LAV approximation is redshift dependent with the drop-off from the full ISW effect in general occurring at lower $\ell$s for higher redshifts. As we are interested here in the dominant, low-$\ell$ part of the ISW effect, the LAV approximation is suitable for our purposes but the reader should be aware that results described for higher $\ell$s in this study are likely to slightly understate the reality of a full non-linear ISW effect. We intend to investigate the expected full $\Lambda$CDM ISW effect from the Jubilee simulation in follow-up studies and we discuss, in \S~\ref{sect:lav_im}, the impact that taking the LAV approximation has on the cross-correlation results.

Figure~\ref{ISW_pow:fig} shows that in our simulation, due to our large box size, we are able to view the ISW effect on very large-scales without an appreciable drop-off in power. This illustrates the requirement, when simulating the ISW, for a box that captures very-large-scale fluctuations in the density field. We show the power spectra of the ISW effect anisotropies in different redshift bins (0$-$1.4, 0$-$0.4, 0.4$-$0.8, 0.8$-$1.2) and compare them to predictions from linear theory. For the $\ell$-range under consideration here the two to correspond closely, as per the findings of \cite{2010MNRAS.407..201C}, who found that the LAV matches linear theory to well past $\ell\sim100$. The power spectra in Figure~\ref{ISW_pow:fig} have been binned. The low-$\ell$ data points ($\ell<6$) are taken in bins of width $\Delta\ell=1$ and show scatter from cosmic variance. As we model a volume with a side-length of $6~h^{-1}\mathrm{Gpc}$ we capture much of the large-scale power in the potential. Despite this, the low-$\ell$ regime of Figure~\ref{ISW_pow:fig} shows that we may be losing a small amount of power on these scales, although not nearly to the extent of that observed in the tiled $1~h^{-1}\mathrm{Gpc}$ box used in \cite{2010MNRAS.407..201C}. This loss in power is made more evident in the comparison between theory and simulation in the last redshift bin ($z=0.809$ to $1.205$) where the largest angular scales (over $6~h^{-1}\mathrm{Gpc}$) that are missing in our simulation box are responsible for the deficit in power at $\ell<5$.

\subsection{LRGs}

\begin{figure*}
  \begin{center}
    \includegraphics[width=6in]{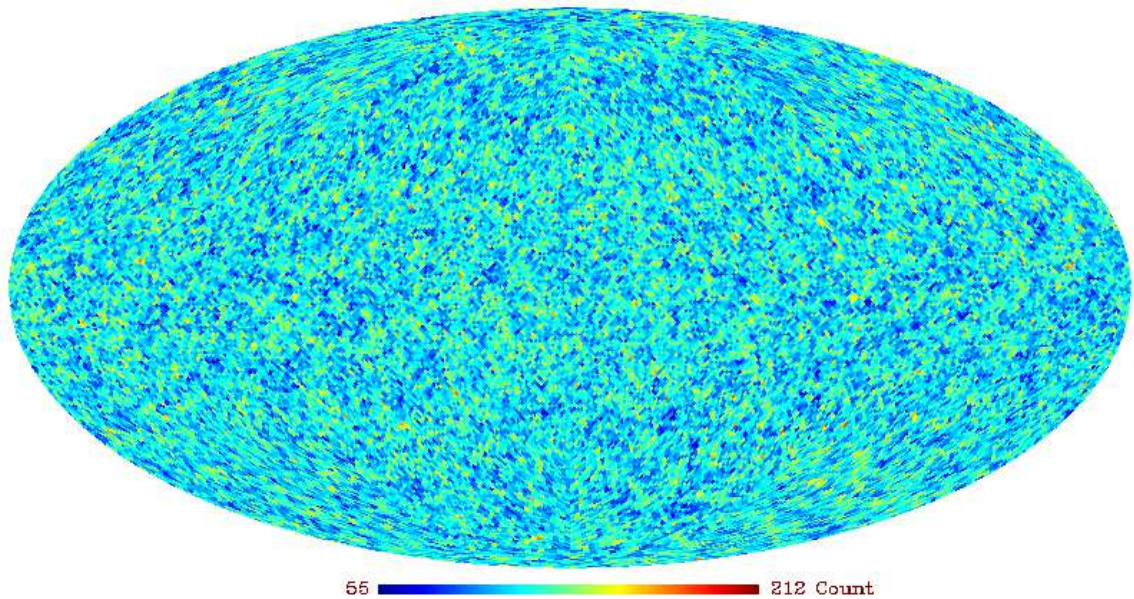}
  \end{center}
  \caption{All-sky projection of the Jubilee mock catalogue LRGs from $z=0$ to $1.4$. A Mollweide projection has been used with $N_{side}=64$. No cuts have been applied to the data.}
    \label{LRG_sky:fig}
\end{figure*}

\begin{figure*}
    \centering
    \begin{subfigure}{3.4in}
	\centering    
        \includegraphics[width=3.4in]{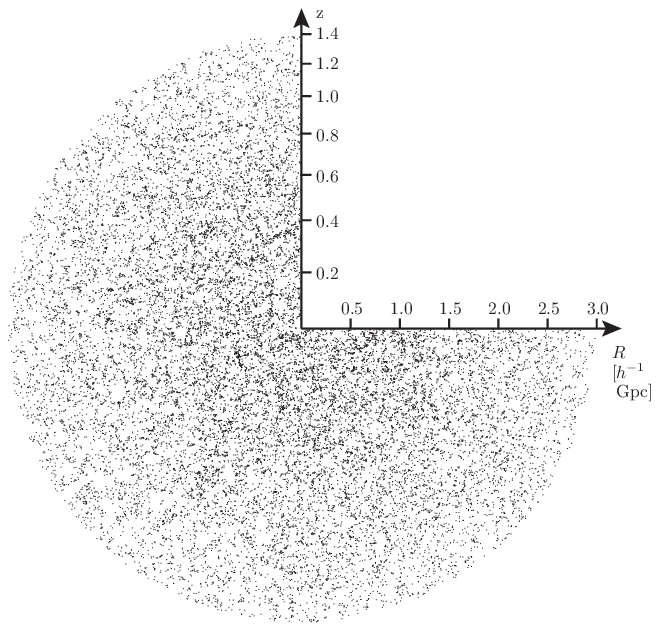}
    \end{subfigure}%
	\quad    
    \begin{subfigure}{3.4in}
	\centering    
        \includegraphics[width=3.4in]{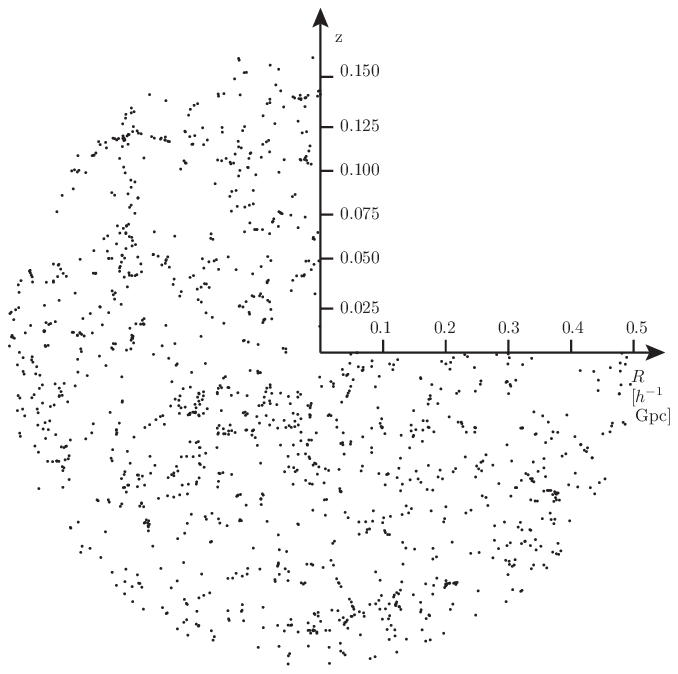}
    \end{subfigure}
    \caption{Projections of the full LRG catalogue distribution from $z = 0$ to $1.4$. \textit{Left panel:} All LRGs lying in a $20~h^{-1}\mathrm{Mpc}$ thick slice, within $3~h^{-1}\mathrm{Gpc}$ of the observer in the centre of the box. \textit{Right panel:} A zoom-in of the local LRG distribution between $0-0.5~h^{-1}\mathrm{Gpc}$ from the observer.}
    \label{survey:fig}
\end{figure*}

In Figure~\ref{LRG_sky:fig} we show a sky map of LRG number counts from all the LRGs in our catalogue between $z=0$ to $1.4$. No cuts of any kind have been applied to this figure and as such it represents the spatial positions on the sky of all the LRGs underneath the black, solid line in Figure~\ref{lrg_hist:fig}. Figure~\ref{survey:fig} shows a projection of the LRGs in the simulation by distance from the observer. Both panels represent a projection that is $20~h^{-1}\mathrm{Mpc}$ deep, with the left-hand panel showing all LRGs out to a radius of $3~h^{-1}\mathrm{Gpc}$ ($z \le 1.4$) and the right-hand panel showing a zoomed-in view of the LRGs out to a radius of $500~h^{-1}\mathrm{Mpc}$ ($z \le 0.17$). Voids and filamentary structures are clearly seen in the distribution. There is little distortion from the peculiar motions of the LRGs. This is due to the fact that the LRGs are all central galaxies and have been assigned the bulk velocity of their host haloes. As such their peculiar velocities are small compared to the higher peculiar velocities of satellite galaxies which orbit the centre of mass of a cluster and create the distinctive `Fingers-of-God' effect.

\begin{figure}
  \begin{center}
    \includegraphics[width=3.4in]{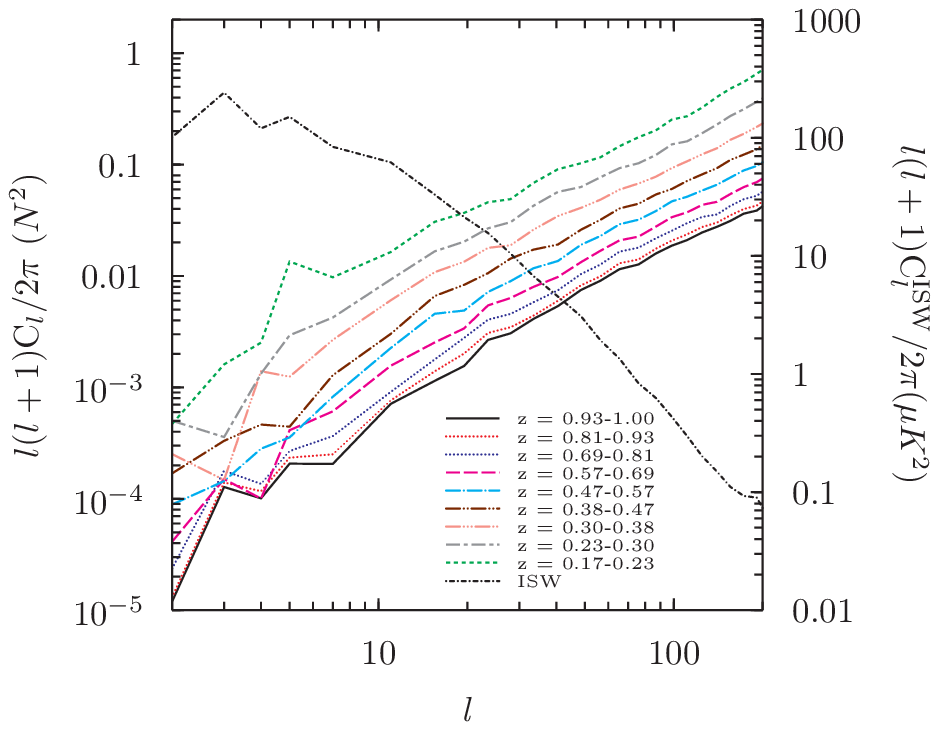}
  \end{center}
  \caption{Power spectra of LRGs for the redshift shells shown in Figure~\ref{ISW_hist:fig}. Overlaid using the second y-axis we also show the ISW power spectrum from Figure~\ref{ISW_pow:fig}.}
    \label{LRG_pow:fig}
\end{figure}

We show the angular power spectrum of our simulated, full-sky catalogues, in Figure~\ref{LRG_pow:fig}. The data has been split into the same redshift shells that we show in Figure~\ref{ISW_hist:fig}. For the purposes of this paper we consider full sky power spectra with no masks which have been corrected for shot noise by removing the expected power from a random, unclustered sample of LRGs. The results in Figure~\ref{LRG_pow:fig} show the expected trend that, as structure formation proceeds, correlations between galaxies grow stronger. We also plot the ISW effect power spectrum on Figure~\ref{LRG_pow:fig} alongside the LRG power.

\subsection{ISW correlation with LSS}
\label{sect:cross}

\begin{figure}
  \begin{center}
    \includegraphics[width=3.2in]{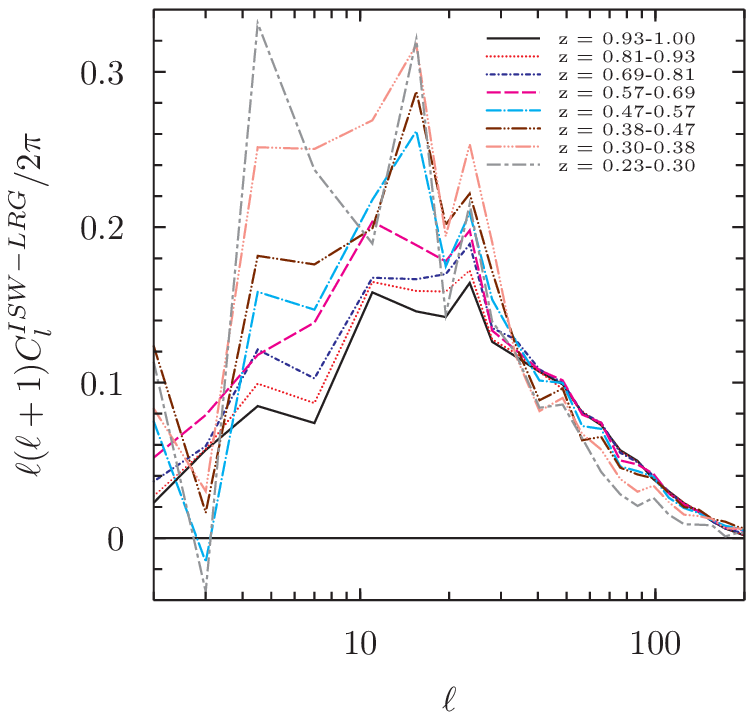}
  \end{center}
  \caption{Cross-correlation signal between the LAV approximation ISW effect integrated between $z = 0$ to $1.4$ and LRGs.}
    \label{cross_z:fig}
\end{figure}

For our cross-correlation analysis we now show how redshift selection of LRGs affects the strength of the ISW-LSS correlation signal. The results and discussion presented here relate to the signal-space for measurements of the ISW-LSS cross-correlation. We stress that this is different from detection-space, in that no signal-to-noise considerations are included in this analysis. We intend to look carefully at results in detection-space in follow-up work. In Figure~\ref{cross_z:fig} we calculate the cross-correlation signal between the ISW effect from $z = 0$ to $1.4$ and LRGs using the same redshift shells as in Figure~\ref{LRG_pow:fig}. The results show that covering the peak of the contribution to the ISW effect, in terms of redshift (i.e.\ $z\sim 0.2$ to $0.5$), is an important factor in producing a strong cross-correlation signal. This result makes no account of LSS survey characteristics, where low values of $\ell$ may not be probed well for a particular survey. Below $\ell \sim 30$ the signal is stronger in the lower redshift bins. Past $\ell \sim 30$ the opposite is true: higher redshift surveys ($z \sim 0.5$ to $1.0$) show a stronger signal.

\subsection{Lensing maps}

\begin{figure*}
    \centering
    \begin{subfigure}{3.3in}
        \includegraphics[width=3.3in]{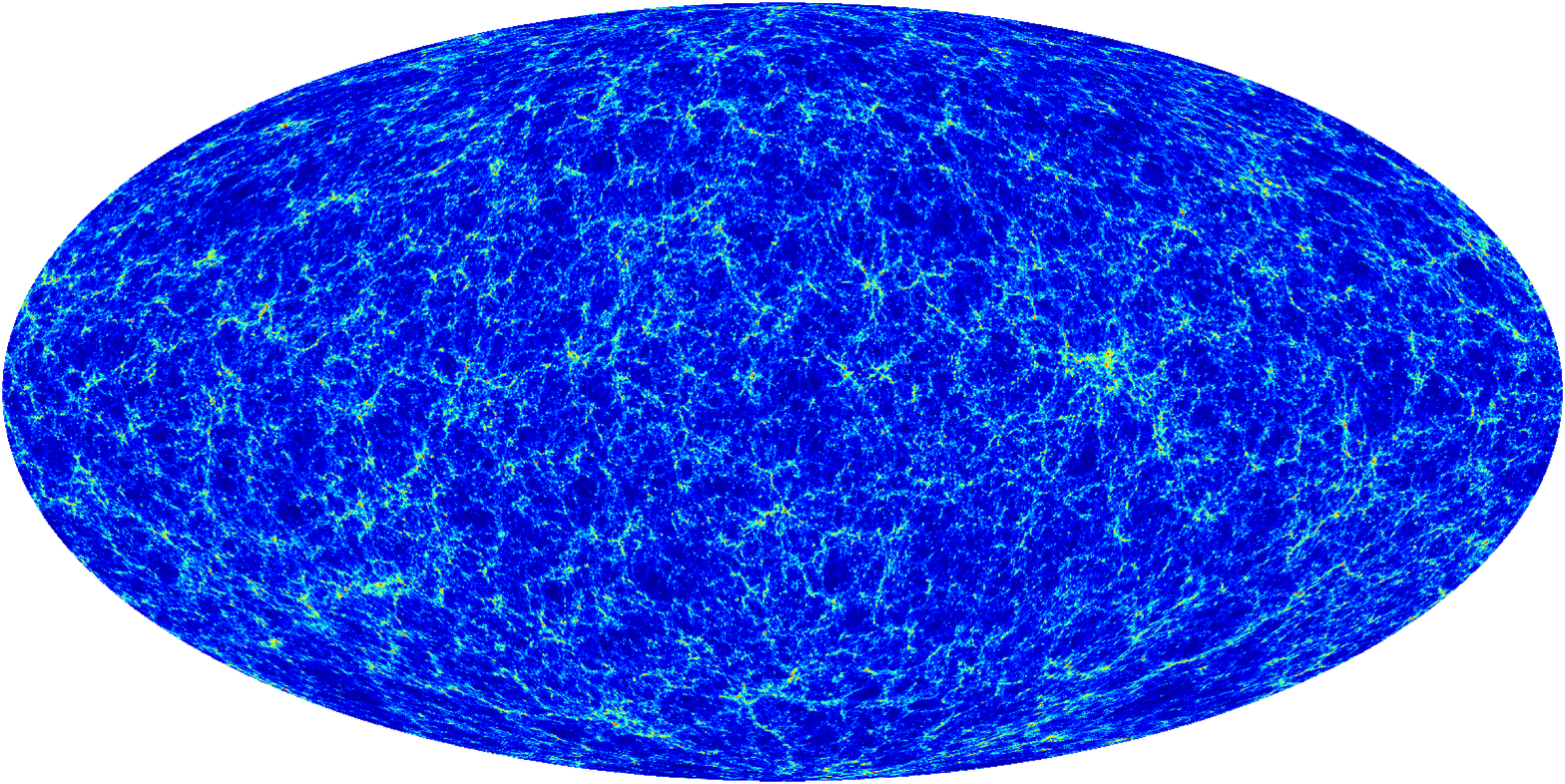}
        \caption{Density field} \label{fig:lens_dens}
    \end{subfigure}%
    \begin{subfigure}{3.3in}
        \includegraphics[width=3.3in]{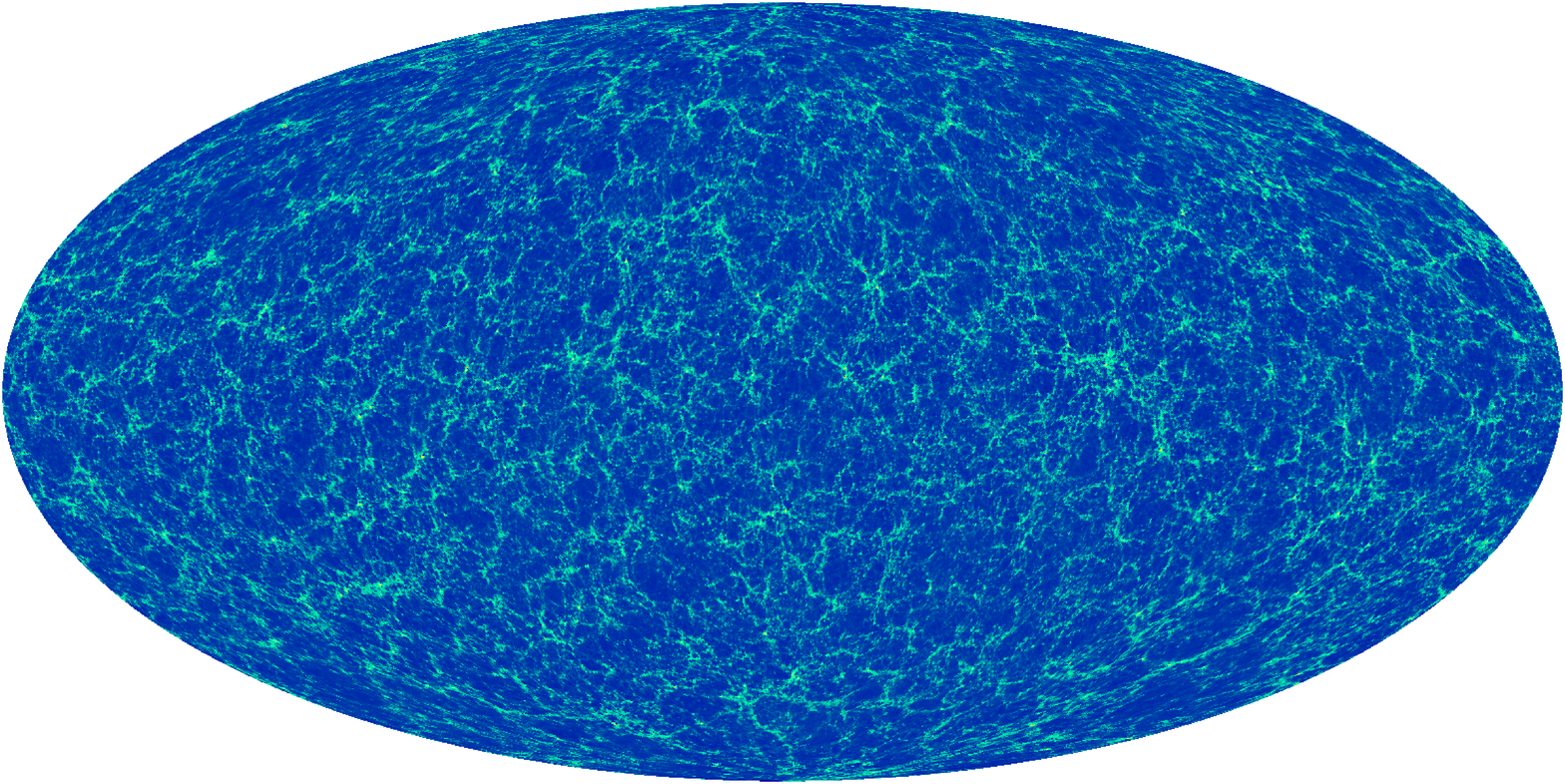}
        \caption{Convergence} \label{fig:lens_con}
    \end{subfigure}%
    \\
    \begin{subfigure}{3.3in}
        \includegraphics[width=3.3in]{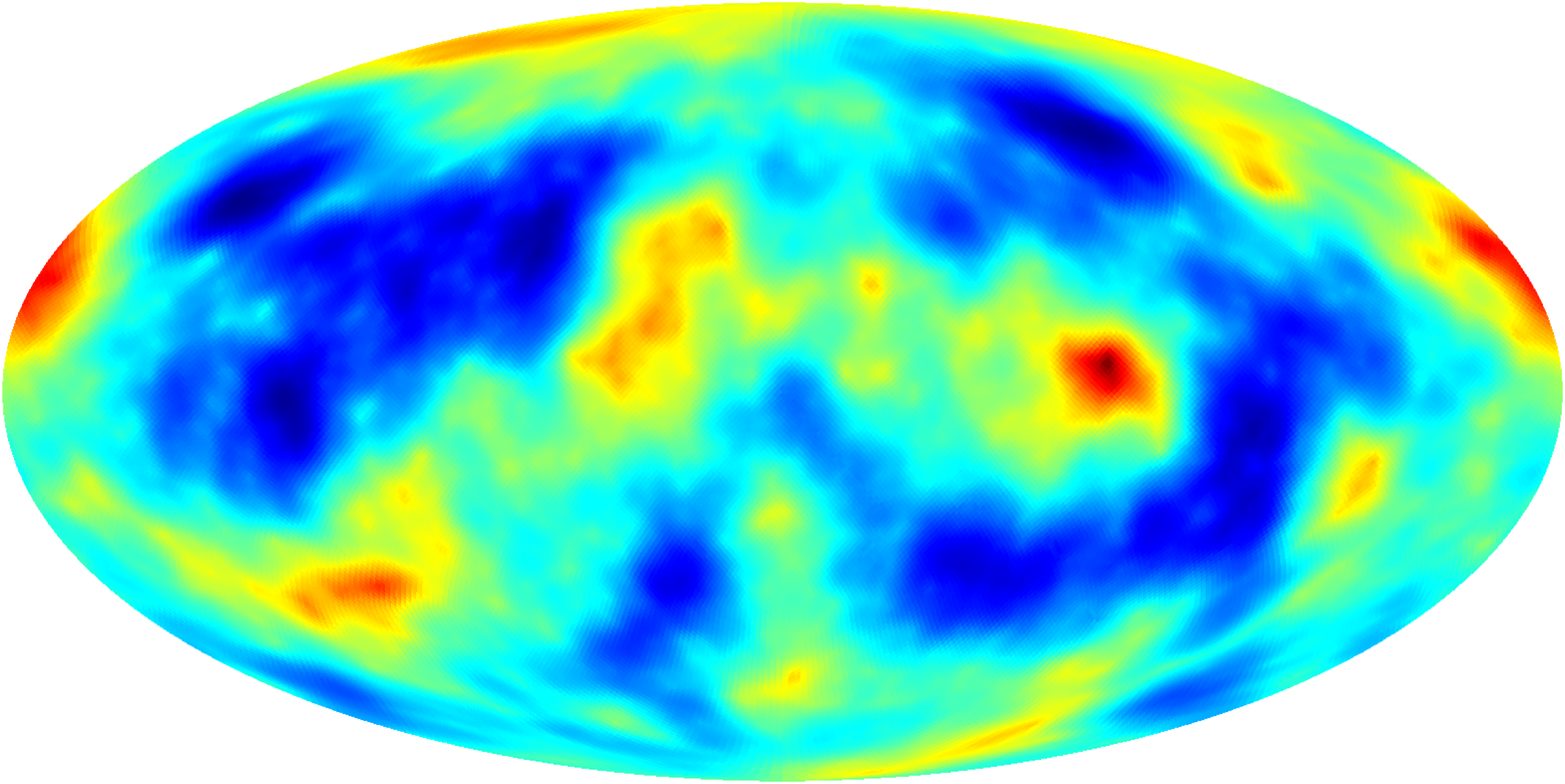}
        \caption{ISW} \label{fig:lens_ISW}
    \end{subfigure}%
    \begin{subfigure}{3.3in}
        \includegraphics[width=3.3in]{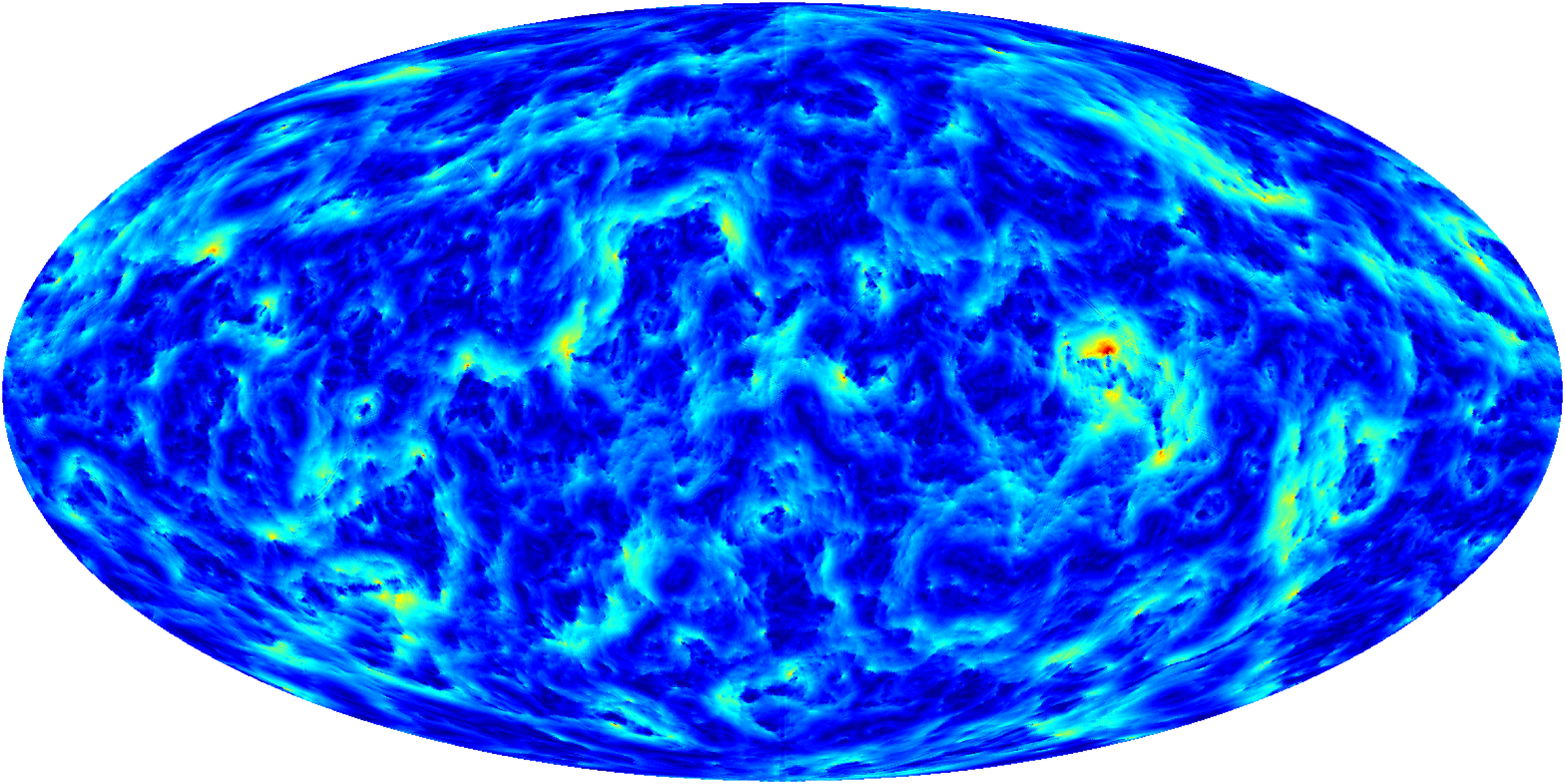} 
        \caption{$\sqrt{\alpha_1^2+\alpha_2^2}$} \label{fig:lens_a_amp}
    \end{subfigure}%
    \\
    \begin{subfigure}{3.3in}
        \includegraphics[width=3.3in]{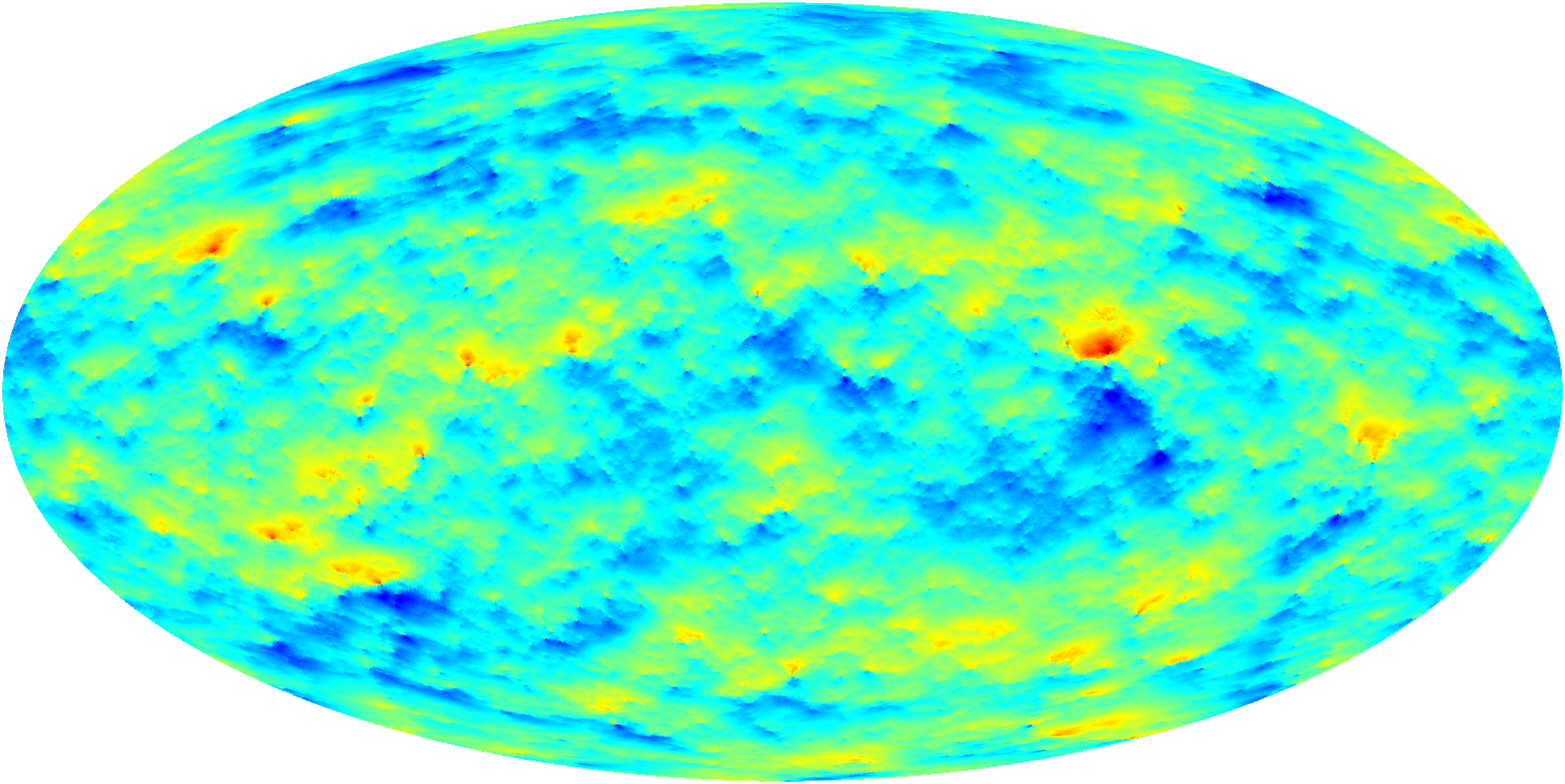}
        \caption{$\alpha_1$} \label{fig:lens_alpha1}
    \end{subfigure}%
    \begin{subfigure}{3.3in}
        \includegraphics[width=3.3in]{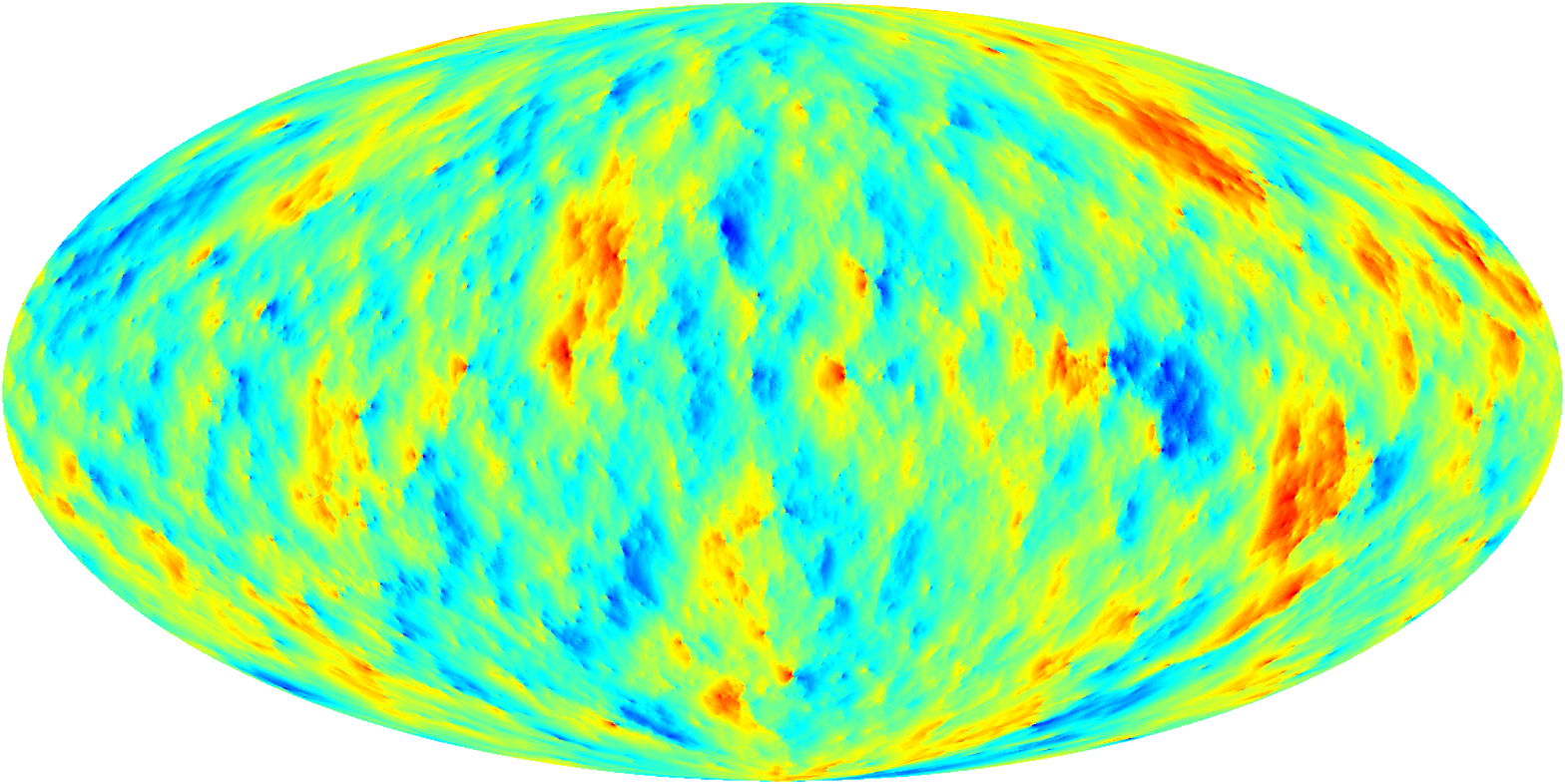}
        \caption{$\alpha_2$} \label{fig:lens_alpha2}
    \end{subfigure}
    \\
    \caption{Complementary output maps from the $z = 0.150$ redshift shell, which spans a redshift range of $z = 0.13$ to $0.17$.}
    \label{fig:lensing}
\end{figure*}

We have produced all-sky lensing maps of the various quantities mentioned in \S~\ref{sect:lensing}, in particular the convergence, $\kappa$, and the deflection angles $\alpha_1$ and $\alpha_2$. We show in Figure~\ref{fig:lensing} a collection of complementary plots for the $z = 0.150$ redshift shell, which spans a redshift range of $z = 0.13$ to $0.17$. The plots show the projected density field (Figure~\ref{fig:lens_dens}), which can be seen closely matches the convergence (Figure~\ref{fig:lens_con}), as expected. We also show the ISW map (Figure~\ref{fig:lens_ISW}) and the effect of a very large overdensity in the right-centre of the plot is very clear, creating as it does a deep potential well and a strong ISW-induced temperature anisotropy. The $\alpha_1$ and $\alpha_2$ plots (Figure~\ref{fig:lens_alpha1} and~\ref{fig:lens_alpha2}) in the region of this large overdensity show characteristic dipoles in the orthogonal $\theta$ and $\phi$ directions (which for this cluster, as it is on the equator of the plot, can be thought of as roughly the same as the up-down and left-right directions respectively). The amplitude of the combined $\alpha_1$ and $\alpha_2$ deflection angles (Figure~\ref{fig:lens_a_amp}) shows very clearly the large overdensity affecting light rays in its region of the sky and features in this map can be seen to correspond to ones in the ISW map.

The results presented here are preliminary, as we intend to do an in-depth analysis of the lensing maps and their relation to the ISW effect and LSS of the simulation in future work.

\section{Discussion}
\label{sect:sum}

\subsection{Implications for signal detection}

Figure~\ref{ISW_hist:fig} shows that, in terms of raw temperature, the maximum impact the ISW effect will have on the photons of the CMB can be expected to occur around $z\sim0.2$ to $0.5$ for a $\Lambda$CDM universe. This immediately implies that a suitable LSS survey should aim to cover the positive peak in this figure, i.e. a redshift range of $z\sim0.1$ to $0.3$. However, Figure~\ref{cross_z:fig} shows that there are other considerations that need to be factored in, specifically relating to the angular dependence of the cross-correlation signal. We see that there is, in fact, a pivot point around $\ell \sim 30$. Below this $\ell$ value lower redshift surveys ($z \sim 0.2$ to $0.5$) are favoured, having characteristic fluctuations on larger angular scales. After $\ell \sim 30$ the opposite is true: higher redshift surveys ($z \sim 0.5$ to $1.0$) show a larger signal, albeit only marginally so when compared to the difference observed in Figure~\ref{cross_z:fig} below $\ell \sim 30$. As we are showing here a full-sky, noise-free case these observations need to be caveated with the fact that, in reality, noise considerations may impact these conclusions (this question will form the basis of some of the follow-up work in the Jubilee-ISW project).

Sky coverage is a major stumbling block in the LSS cross-correlation approach because anything less than a full sky survey begins to impact the signal-to-noise level of the detection. This places a strain on any LSS survey -- which typically have to balance sky coverage versus survey depth -- that aims to optimise an ISW measurement. A detailed analysis of the various signal-to-noise considerations in the LSS-ISW correlation measurement can be found in \cite{2007MNRAS.381.1347C}. Previous correlation measurements have utilised a variety of LSS catalogues including the NVSS \citep{1998AJ....115.1693C} radio survey, which had a sky coverage of 82\%, and the SDSS galaxy survey with a sky coverage of 35\% \citep{2012ApJS..203...21A}. These surveys have their own pros and cons. Whilst the NVSS has excellent sky coverage it has only $\sim 1.4$ million objects and these are found across a wide redshift distribution \citep[$z = 0$ to $2+$, we refer the reader to Figure 2 from][]{2013arXiv1303.5079P}. The SDSS, on the other hand, contains many sources (almost 1 billion galaxies in total), in particular LRGs, across a redshift range \citep[$z\sim0$ to $0.8$, see Figure 2 of][]{2013arXiv1303.5079P} that is very well suited to ISW effect detection. However, because of its lesser sky coverage, it has a high noise level on larger scales. This implies that, unless a survey is able to probe $\ell < 30$ scales, a redshift range of $z \sim 0.6$ to $1.0$ is more suitable for ISW detection efforts.

There are future surveys that will have appropriate sky and redshift footprints, in particular the HI Evolutionary Map of the Universe (EMU) survey \citep{2011JApA...32..599N}, that will be performed using the ASKAP telescope. EMU is a pathfinder for the Square Kilometre Array (SKA), and will detect sources across a broad range of redshifts, $z\sim0-6$, in particular low-redshift star-forming galaxies at $z<2$. Its sky coverage will be roughly the same as for the NVSS and the intention is for its data to be combined with another HI survey, Westerbork Observations of the Deep APERTIF Northern sky (WODAN) \citep{2011JApA...32..557R}, which will cover the remaining patch of the Northern Hemisphere that EMU cannot see.

\subsection{Model discrimination using the ISW}

The hope that the ISW signal can in the future be used to help discriminate between cosmological models depends on our ability to measure the signal accurately. Results from stacking approaches are currently placing the $\Lambda$CDM model under scrutiny. Work by previous authors on cross-correlations between the CMB and the ISW have attempted to constrain cosmological parameters based on the observed correlations \citep{2005PhRvD..72d3525P,2006PhRvD..74d3524P,2006MNRAS.365..171G,2008PhRvD..77l3520G}. These results are summarised in \cite{2013arXiv1303.5079P}, with the general consensus being that ISW observations have constrained $\Omega_\Lambda$ to a value of $\Omega_\Lambda \approx 0.75 \pm 20\%$; $\Omega_K$ to be flat to within a few percent and the equation of state parameter to be $\omega \approx 1$ with no strong evolution. These results highlight the fact that the ISW effect does not constrain the $\Lambda$CDM model to anything like the precision of the standard datasets (CMB and BAOs). However, for a universe containing an amount of warm dark matter or one with a temporally varying dark energy component, the ISW effect should be an aid in constraining the models we use to describe them.

For alternative cosmological models a variety of expectations of ISW signal arise. A study by \cite{2012ApJ...744....3M} on the effect of massive neutrinos on the ISW-LSS correlation signal, along with the expectations of different coupled dark energy models, shows that model discrimination typically involves a difference in the expected height of the peak in the cross-correlations (using the cross-correlation multiplied by $\ell(\ell+1)$ as we do in this paper). They also note that the models are better discriminated between at higher redshifts. As redshift selection cuts modulate both the peak height and possibly the peak position of the cross-correlation signal (Figure~\ref{cross_z:fig}), redshift selection effects need to be carefully deciphered. The Jubilee ISW project will help determine the best strategies to discriminate among models since we will provide the tools (ISW maps and associated catalogs) that will make possible the validation and calibration of new techniques against simulated data.

\subsection{Impact of using the LAV approximation}
\label{sect:lav_im}

This study has focused on low-$\ell$ results based on the LAV approximation. As such, results for higher $\ell$ values ($\ell > 100$) presented here will deviate significantly from the expected ISW-induced anisotropies which include velocity information. Studies by \cite{2009MNRAS.396..772C} and \cite{2010MNRAS.407..201C} have looked in detail at the specific contribution the velocity information makes to the ISW anisotropies. Figure 2 in \cite{2009MNRAS.396..772C} summarises the expected deviation of the full result from that of linear theory. Essentially, past an $\ell \sim 60$ for low redshifts ($z < 0.5$), the deviation from the full ISW anisotropies begins to become significant. This evolves to lower $\ell$ for higher redshifts until at $z \sim 1$ the deviation from linear theory begins to become significant at around $\ell \sim 40-50$. The LAV approximation, which uses full, simulated information from the density field but combines it with a linear theory velocity prescription, follows the linear theory prediction very closely at the $\ell$ values where the non-linear contribution becomes significant.

The effect of the non-linear component on the cross-correlation with ISW anisotropies is to suppress correlation at $\ell$ values that are much higher than where the non-linearities become significant in terms of raw power. \cite{2009MNRAS.396..772C} found that the deviation from the expected linear CMB-LSS cross-correlation signal only became significant at $\ell \gtrsim 500$, which implies that all of the cross-correlation results presented here can be taken as accurate.

\subsection{Future work}
 
We intend to undertake a range of follow-up studies into the ISW effect in the Jubilee simulation. In particular we will use synthetic CMB maps to study the recovery of the ISW effect using LSS cross-correlation, stacking and lensing methodologies. We will be creating a mock NVSS-like catalogue to investigate the expected signal from broad-sky radio surveys as well as extending the redshift range of our ISW calculation and also calculating a full non-linear ISW effect. 

In future work we intend to examine the expected $\Lambda$CDM signal from a stacking analysis designed to mimic the measurement in \citet*{2008ApJ...683L..99G}. This will involve applying the void and structure finding algorithms ZOBOV and VOBOZ\footnote{\url{http://skysrv.pha.jhu.edu/~neyrinck/voboz/}} \citep*{2005MNRAS.356.1222N} to our sample LRG catalogues and then stacking images of the CMB along the lines of sight of structures found by these algorithms.
 
\section{Acknowledgements}
 
We thank Zheng Zheng for helpful correspondence regarding the HOD model used in this work. The simulation was performed on the Juropa supercomputer of the J\"ulich Supercomputing Centre (JSC). Some of the results in this paper have been derived using the {\small HEALPix} package. WW thanks The Southeast Physics Network (SEPNet) for providing funding for his research. ITI was supported by The SEPNet and the Science and Technology Facilities Council grants ST/F002858/1 and ST/I000976/1. AK is supported by the Spanish Ministerio de Ciencia e Innovacion (MICINN) in Spain through the Ramon y Cajal programme as well as the grants AYA2009-13875-C03-02, AYA2009-12792-C03-03, AYA2012-31101, CSD2009-00064, and CAM S2009/ESP-1496. GY acknowledges support from MINECO (Spain) under research grants AYA2009-13875-C03-02, AYA2012-31101, FPA2009-08958 and Consolider Ingenio SyeC CSD2007-0050. SH acknowledges support from the Academy of Finland grant 131454. SN acknowledges support from the Sofja Kovalevskaja program of the Alexander von Humboldt foundation. JMD, RBB, JGN, EMG, PV thank financial support form AYA2010-21766- C03-01, AYA2012-39475-C02-01 and Consolider-Ingenio 2010 CSD2010-00064.

\bibliographystyle{mn}

\begin{thebibliography}{73}
\providecommand{\natexlab}[1]{#1}

\bibitem[{{Afshordi} et~al.(2004){Afshordi}, {Loh} \&
  {Strauss}}]{2004PhRvD..69h3524A}
{Afshordi} N., {Loh} Y.~S., {Strauss} M.~A., 2004, \prd, 69, 083524

\bibitem[{{Ahn} et~al.(2012)}]{2012ApJS..203...21A}
{Ahn} C.~P. et~al., 2012, \apjs, 203, 21

\bibitem[{{Amiaux} et~al.(2012)}]{2012SPIE.8442E..0ZA}
{Amiaux} J. et~al., 2012, in Society of Photo-Optical Instrumentation Engineers
  (SPIE) Conference Series. Vol. 8442

\bibitem[{{Barber} et~al.(1999){Barber}, {Thomas} \&
  {Couchman}}]{1999MNRAS.310..453B}
{Barber} A.~J., {Thomas} P.~A., {Couchman} H.~M.~P., 1999, \mnras, 310, 453

\bibitem[{{Behroozi} et~al.(2013){Behroozi}, {Wechsler} \&
  {Wu}}]{2013ApJ...762..109B}
{Behroozi} P.~S., {Wechsler} R.~H., {Wu} H.~Y., 2013, \apj, 762, 109

\bibitem[{{Boughn} \& {Crittenden}(2004)}]{2004Natur.427...45B}
{Boughn} S., {Crittenden} R., 2004, \nat, 427, 45

\bibitem[{{Boughn} \& {Crittenden}(2002)}]{2002PhRvL..88b1302B}
{Boughn} S.~P., {Crittenden} R.~G., 2002, Physical Review Letters, 88, 021302

\bibitem[{{Cabr{\'e}} et~al.(2007){Cabr{\'e}}, {Fosalba}, {Gazta{\~n}aga} \&
  {Manera}}]{2007MNRAS.381.1347C}
{Cabr{\'e}} A., {Fosalba} P., {Gazta{\~n}aga} E., {Manera} M., 2007, \mnras,
  381, 1347

\bibitem[{{Cai} et~al.(2009){Cai}, {Cole}, {Jenkins} \&
  {Frenk}}]{2009MNRAS.396..772C}
{Cai} Y.~C., {Cole} S., {Jenkins} A., {Frenk} C., 2009, \mnras, 396, 772

\bibitem[{{Cai} et~al.(2010){Cai}, {Cole}, {Jenkins} \&
  {Frenk}}]{2010MNRAS.407..201C}
{Cai} Y.~C., {Cole} S., {Jenkins} A., {Frenk} C.~S., 2010, \mnras, 407, 201

\bibitem[{{Carbone} et~al.(2008){Carbone}, {Springel}, {Baccigalupi},
  {Bartelmann} \& {Matarrese}}]{2008MNRAS.388.1618C}
{Carbone} C., {Springel} V., {Baccigalupi} C., {Bartelmann} M., {Matarrese} S.,
  2008, \mnras, 388, 1618

\bibitem[{{Carbone} et~al.(2013){Carbone}, {Baldi}, {Pettorino} \&
  {Baccigalupi}}]{Carbone2013}
{Carbone} C., {Baldi} M., {Pettorino} V., {Baccigalupi} C., 2013,
  arXiv:1305.0829

\bibitem[{{Condon} et~al.(1998){Condon}, {Cotton}, {Greisen}, {Yin}, {Perley},
  {Taylor} \& {Broderick}}]{1998AJ....115.1693C}
{Condon} J.~J., {Cotton} W.~D., {Greisen} E.~W., {Yin} Q.~F., {Perley} R.~A.,
  {Taylor} G.~B., {Broderick} J.~J., 1998, \aj, 115, 1693

\bibitem[{{Cooray}(2002)}]{2002PhRvD..65j3510C}
{Cooray} A., 2002, \prd, 65, 103510

\bibitem[{{Crittenden} \& {Turok}(1996)}]{1996PhRvL..76..575C}
{Crittenden} R.~G., {Turok} N., 1996, Physical Review Letters, 76, 575

\bibitem[{{Das} \& {Bode}(2008)}]{2008ApJ...682....1D}
{Das} S., {Bode} P., 2008, \apj, 682, 1

\bibitem[{{Douspis} et~al.(2008){Douspis}, {Castro}, {Caprini} \&
  {Aghanim}}]{2008A&A...485..395D}
{Douspis} M., {Castro} P.~G., {Caprini} C., {Aghanim} N., 2008, \aap, 485, 395

\bibitem[{Dunkley et~al.(2009)}]{Dunkley:2008ie}
Dunkley J. et~al., 2009, \apjs, 180, 306

\bibitem[{{Eisenstein} et~al.(2005)}]{2005ApJ...633..560E}
{Eisenstein} D.~J. et~al., 2005, \apj, 633, 560

\bibitem[{{Flender} et~al.(2013){Flender}, {Hotchkiss} \&
  {Nadathur}}]{2013JCAP...02..013F}
{Flender} S., {Hotchkiss} S., {Nadathur} S., 2013, JCAP, 2, 013

\bibitem[{{Fosalba} et~al.(2003){Fosalba}, {Gazta{\~n}aga} \&
  {Castander}}]{2003ApJ...597L..89F}
{Fosalba} P., {Gazta{\~n}aga} E., {Castander} F.~J., 2003, \apjl, 597, L89

\bibitem[{{Fosalba} et~al.(2008){Fosalba}, {Gazta{\~n}aga}, {Castander} \&
  {Manera}}]{2008MNRAS.391..435F}
{Fosalba} P., {Gazta{\~n}aga} E., {Castander} F.~J., {Manera} M., 2008, \mnras,
  391, 435

\bibitem[{{Gazta{\~n}aga} et~al.(2006){Gazta{\~n}aga}, {Manera} \&
  {Multam{\"a}ki}}]{2006MNRAS.365..171G}
{Gazta{\~n}aga} E., {Manera} M., {Multam{\"a}ki} T., 2006, \mnras, 365, 171

\bibitem[{{Giannantonio} et~al.(2008){Giannantonio}, {Scranton}, {Crittenden},
  {Nichol}, {Boughn}, {Myers} \& {Richards}}]{2008PhRvD..77l3520G}
{Giannantonio} T., {Scranton} R., {Crittenden} R.~G., {Nichol} R.~C., {Boughn}
  S.~P., {Myers} A.~D., {Richards} G.~T., 2008, \prd, 77, 123520

\bibitem[{Gill et~al.(2004)Gill, Knebe \& Gibson}]{Gill:2004km}
Gill S.~P., Knebe A., Gibson B.~K., 2004, \MNRAS, 351, 399

\bibitem[{{G{\'o}rski} et~al.(2005){G{\'o}rski}, {Hivon}, {Banday}, {Wandelt},
  {Hansen}, {Reinecke} \& {Bartelmann}}]{2005ApJ...622..759G}
{G{\'o}rski} K.~M., {Hivon} E., {Banday} A.~J., {Wandelt} B.~D., {Hansen}
  F.~K., {Reinecke} M., {Bartelmann} M., 2005, \apj, 622, 759

\bibitem[{{Granett} et~al.(2008){Granett}, {Neyrinck} \&
  {Szapudi}}]{2008ApJ...683L..99G}
{Granett} B.~R., {Neyrinck} M.~C., {Szapudi} I., 2008, \apjl, 683, L99

\bibitem[{Harnois-Deraps et~al.(2012)Harnois-Deraps, Pen, Iliev, Merz, Emberson
  et~al.}]{HarnoisDeraps:2012he}
Harnois-Deraps J., Pen U.~L., Iliev I.~T., Merz H., Emberson J. et~al., 2012,
  arXiv:1208.5098

\bibitem[{{Heath}(1977)}]{1977MNRAS.179..351H}
{Heath} D.~J., 1977, \mnras, 179, 351

\bibitem[{{Hernandez-Monteagudo} \& {Smith}(2012)}]{2012arXiv1212.1174H}
{Hernandez-Monteagudo} C., {Smith} R.~E., 2012, arXiv:1212.1174

\bibitem[{{Hilbert} et~al.(2009){Hilbert}, {Hartlap}, {White} \&
  {Schneider}}]{2009A&A...499...31H}
{Hilbert} S., {Hartlap} J., {White} S.~D.~M., {Schneider} P., 2009, \aap, 499,
  31

\bibitem[{Hockney \& Eastwood(1988)}]{Hockney:1988:CSU}
Hockney R.~W., Eastwood J.~W., 1988, Computer Simulation Using Particles. Adam
  Hilger Ltd., Bristol, UK

\bibitem[{{Hoekstra} \& {Jain}(2008)}]{2008ARNPS..58...99H}
{Hoekstra} H., {Jain} B., 2008, Annual Review of Nuclear and Particle Science,
  58, 99

\bibitem[{{Hu} \& {Sugiyama}(1994)}]{1994PhRvD..50..627H}
{Hu} W., {Sugiyama} N., 1994, \prd, 50, 627

\bibitem[{Hunt \& Sarkar(2010)}]{Hunt:2008wp}
Hunt P., Sarkar S., 2010, Mon.Not.Roy.Astron.Soc., 401, 547

\bibitem[{{Ilic} et~al.(2013){Ilic}, {Langer} \&
  {Douspis}}]{2013arXiv1301.5849I}
{Ilic} S., {Langer} M., {Douspis} M., 2013, arXiv:1301.5849

\bibitem[{{Jain} et~al.(2000){Jain}, {Seljak} \& {White}}]{2000ApJ...530..547J}
{Jain} B., {Seljak} U., {White} S., 2000, \apj, 530, 547

\bibitem[{{Kamionkowski} \& {Spergel}(1994)}]{1994ApJ...432....7K}
{Kamionkowski} M., {Spergel} D.~N., 1994, \apj, 432, 7

\bibitem[{{Kazin} et~al.(2010)}]{2010ApJ...710.1444K}
{Kazin} E.~A. et~al., 2010, \apj, 710, 1444

\bibitem[{{Kiessling} et~al.(2011){Kiessling}, {Heavens}, {Taylor} \&
  {Joachimi}}]{2011MNRAS.414.2235K}
{Kiessling} A., {Heavens} A.~F., {Taylor} A.~N., {Joachimi} B., 2011, \mnras,
  414, 2235

\bibitem[{Knollmann \& Knebe(2009)}]{Knollmann:2009pb}
Knollmann S.~R., Knebe A., 2009, \apjs, 182, 608

\bibitem[{Komatsu et~al.(2009)}]{Komatsu:2008hk}
Komatsu E. et~al., 2009, \apjs, 180, 330

\bibitem[{{Lacey} \& {Cole}(1994)}]{1994MNRAS.271..676L}
{Lacey} C., {Cole} S., 1994, \MNRAS, 271, 676

\bibitem[{{Lawrence} et~al.(2010){Lawrence}, {Heitmann}, {White}, {Higdon},
  {Wagner}, {Habib} \& {Williams}}]{2010ApJ...713.1322L}
{Lawrence} E., {Heitmann} K., {White} M., {Higdon} D., {Wagner} C., {Habib} S.,
  {Williams} B., 2010, \apj, 713, 1322

\bibitem[{Lewis et~al.(2000)Lewis, Challinor \& Lasenby}]{Lewis:1999bs}
Lewis A., Challinor A., Lasenby A., 2000, Astrophys. J., 538, 473

\bibitem[{{Mainini} \& {Mota}(2012)}]{2012ApJ...744....3M}
{Mainini} R., {Mota} D.~F., 2012, \apj, 744, 3

\bibitem[{{Mandelbaum} et~al.(2013)}]{2013MNRAS.tmp.1228M}
{Mandelbaum} R., {Slosar} A., {Baldauf} T., {Seljak} U., {Hirata} C.~M.,
  {Nakajima} R., {Reyes} R., {Smith} R.~E., 2013, \mnras

\bibitem[{{McEwen} et~al.(2007){McEwen}, {Vielva}, {Hobson},
  {Mart{\'{\i}}nez-Gonz{\'a}lez} \& {Lasenby}}]{2007MNRAS.376.1211M}
{McEwen} J.~D., {Vielva} P., {Hobson} M.~P., {Mart{\'{\i}}nez-Gonz{\'a}lez} E.,
  {Lasenby} A.~N., 2007, \mnras, 376, 1211

\bibitem[{{Nadathur} et~al.(2012){Nadathur}, {Hotchkiss} \&
  {Sarkar}}]{2012JCAP...06..042N}
{Nadathur} S., {Hotchkiss} S., {Sarkar} S., 2012, \jcap, 6, 042

\bibitem[{{Neyrinck} et~al.(2005){Neyrinck}, {Gnedin} \&
  {Hamilton}}]{2005MNRAS.356.1222N}
{Neyrinck} M.~C., {Gnedin} N.~Y., {Hamilton} A.~J.~S., 2005, \mnras, 356, 1222

\bibitem[{{Nolta} et~al.(2004)}]{2004ApJ...608...10N}
{Nolta} M.~R. et~al., 2004, \apj, 608, 10

\bibitem[{{Norris}(2011)}]{2011JApA...32..599N}
{Norris} R.~P., 2011, Journal of Astrophysics and Astronomy, 32, 599

\bibitem[{{Padmanabhan} et~al.(2005){Padmanabhan}, {Hirata}, {Seljak},
  {Schlegel}, {Brinkmann} \& {Schneider}}]{2005PhRvD..72d3525P}
{Padmanabhan} N., {Hirata} C.~M., {Seljak} U., {Schlegel} D.~J., {Brinkmann}
  J., {Schneider} D.~P., 2005, \prd, 72, 043525

\bibitem[{{Pietrobon} et~al.(2006){Pietrobon}, {Balbi} \&
  {Marinucci}}]{2006PhRvD..74d3524P}
{Pietrobon} D., {Balbi} A., {Marinucci} D., 2006, \prd, 74, 043524

\bibitem[{{Planck Collaboration}(2013{\natexlab{a}})}]{2013arXiv1303.5062P}
{Planck Collaboration}, 2013{\natexlab{a}}, arXiv:1303.5062

\bibitem[{{Planck Collaboration}(2013{\natexlab{b}})}]{2013arXiv1303.5076P}
{Planck Collaboration}, 2013{\natexlab{b}}, arXiv:1303.5076

\bibitem[{{Planck Collaboration}(2013{\natexlab{c}})}]{2013arXiv1303.5079P}
{Planck Collaboration}, 2013{\natexlab{c}}, arXiv:1303.5079

\bibitem[{{Planck Collaboration}(2013{\natexlab{d}})}]{2013arXiv1303.5084P}
{Planck Collaboration}, 2013{\natexlab{d}}, arXiv:1303.5084

\bibitem[{{Planck Collaboration}(2013{\natexlab{e}})}]{2013arXiv1303.5077P}
{Planck Collaboration}, 2013{\natexlab{e}}, arXiv:1303.5077

\bibitem[{{Rees} \& {Sciama}(1968)}]{1968Natur.217..511R}
{Rees} M.~J., {Sciama} D.~W., 1968, \nat, 217, 511

\bibitem[{{R{\"o}ttgering} et~al.(2011)}]{2011JApA...32..557R}
{R{\"o}ttgering} H. et~al., 2011, Journal of Astrophysics and Astronomy, 32,
  557

\bibitem[{{Sachs} \& {Wolfe}(1967)}]{1967ApJ...147...73S}
{Sachs} R.~K., {Wolfe} A.~M., 1967, \apj, 147, 73

\bibitem[{{Schiavon} et~al.(2012)}]{2012MNRAS.427.3044S}
{Schiavon} F., {Finelli} F., {Gruppuso} A., {Marcos-Caballero} A., {Vielva} P.,
  {Crittenden} R.~G., {Barreiro} R.~B., {Mart{\'{\i}}nez-Gonz{\'a}lez} E.,
  2012, \mnras, 427, 3044

\bibitem[{{Teyssier} et~al.(2009)}]{2009A&A...497..335T}
{Teyssier} R. et~al., 2009, \aap, 497, 335

\bibitem[{{Vale} \& {White}(2003)}]{2003ApJ...592..699V}
{Vale} C., {White} M., 2003, \apj, 592, 699

\bibitem[{{Vielva} et~al.(2006){Vielva}, {Mart{\'{\i}}nez-Gonz{\'a}lez} \&
  {Tucci}}]{2006MNRAS.365..891V}
{Vielva} P., {Mart{\'{\i}}nez-Gonz{\'a}lez} E., {Tucci} M., 2006, \mnras, 365,
  891

\bibitem[{{Watson} et~al.(2012){Watson}, {Iliev}, {D'Aloisio}, {Knebe},
  {Shapiro} \& {Yepes}}]{2012arXiv1212.0095W}
{Watson} W.~A., {Iliev} I.~T., {D'Aloisio} A., {Knebe} A., {Shapiro} P.~R.,
  {Yepes} G., 2012, arXiv:1212.0095

\bibitem[{{Watson} et~al.(2013){Watson}, {Iliev}, {Diego}, {Gottl{\"o}ber},
  {Knebe}, {Mart{\'{\i}}nez-Gonz{\'a}lez} \& {Yepes}}]{2013arXiv1305.1976W}
{Watson} W.~A., {Iliev} I.~T., {Diego} J.~M., {Gottl{\"o}ber} S., {Knebe} A.,
  {Mart{\'{\i}}nez-Gonz{\'a}lez} E., {Yepes} G., 2013, arXiv:1305.1976

\bibitem[{{Wen} et~al.(2012){Wen}, {Han} \& {Liu}}]{2012ApJS..199...34W}
{Wen} Z.~L., {Han} J.~L., {Liu} F.~S., 2012, \apjs, 199, 34

\bibitem[{{Zel'dovich}(1970)}]{1970A&A.....5...84Z}
{Zel'dovich} Y.~B., 1970, \aap, 5, 84

\bibitem[{{Zheng} et~al.(2009){Zheng}, {Zehavi}, {Eisenstein}, {Weinberg} \&
  {Jing}}]{2009ApJ...707..554Z}
{Zheng} Z., {Zehavi} I., {Eisenstein} D.~J., {Weinberg} D.~H., {Jing} Y.~P.,
  2009, \apj, 707, 554

\bibitem[{{Zheng} et~al.(2005)}]{2005ApJ...633..791Z}
{Zheng} Z. et~al., 2005, \apj, 633, 791

\bibitem[{{Zitrin} et~al.(2012){Zitrin}, {Bartelmann}, {Umetsu}, {Oguri} \&
  {Broadhurst}}]{2012MNRAS.426.2944Z}
{Zitrin} A., {Bartelmann} M., {Umetsu} K., {Oguri} M., {Broadhurst} T., 2012,
  \mnras, 426, 2944

\end{thebibliography}

\end{document}